\documentclass[twocolumn,tighten,times,tracking,twocolappendix]{aastex}

\usepackage{amsmath}
\usepackage{cancel}
\usepackage{color}
\usepackage{longtable}
\usepackage{ulem,array}
\usepackage{grffile}
\usepackage{comment}
\usepackage{hyperref}
\hypersetup{
pageanchor=false,
colorlinks=true,
linkcolor=blue,
citecolor=blue
}

\usepackage{booktabs}   

\newcommand{\ODM}{\Omega_\mathrm{DM}}
\newcommand{\OB}{\Omega_\mathrm{b}}

\usepackage{multirow}
\usepackage{amssymb}
\makeatletter
\newcommand{\xRightarrow}[2][]{\ext@arrow 0359\Rightarrowfill@{#1}{#2}}
\makeatother

\usepackage[mathscr]{euscript}

\DeclareSymbolFont{matha}{OML}{txmi}{m}{it}
\DeclareMathSymbol{\varv}{\mathord}{matha}{29}

\SetSymbolFont{symbols}{bold}{OMS}{cmsy}{b}{n}
\DeclareSymbolFont{bmisymbols}{OML}{cmm}{b}{it}
\DeclareMathSymbol{\bvarv}{0}{bmisymbols}{"1D}

\definecolor{ForestGreen}{rgb}{0.133,0.545,0.133}

\begin{document}

\author[0000-0001-9625-5929]{Mohamad Shalaby}
\affiliation{Perimeter Institute for Theoretical Physics, 31 Caroline Street North, Waterloo, ON, N2L 2Y5, Canada}
\affiliation{Waterloo Centre for Astrophysics, University of Waterloo, Waterloo, ON N2L 3G1, Canada}
\affiliation{Department of Physics and Astronomy, University of Waterloo, 200 University Avenue West, Waterloo, ON, N2L 3G1, Canada}
\affiliation{%
Horizon AstroPhysics Initiative (HAPI) Fellow}

\author[0000-0002-3351-760X]{Avery Broderick}
\affiliation{Perimeter Institute for Theoretical Physics, 31 Caroline Street North, Waterloo, ON, N2L 2Y5, Canada}
\affiliation{Waterloo Centre for Astrophysics, University of Waterloo, Waterloo, ON N2L 3G1, Canada}
\affiliation{Department of Physics and Astronomy, University of Waterloo, 200 University Avenue West, Waterloo, ON, N2L 3G1, Canada}

\title{%
The Sound of the Universe: A Resonant Gravitational Instability Driven by Baryon-Dark Matter Relative Drift
}
\shorttitle{%
The Sound of the Universe%
}
\shortauthors{Shalaby \& Broderick}

\begin{abstract}
Dark matter and baryons acquire a relative velocity after decoupling in the early Universe. Baryons are gravitationally unstable only above their Jeans scale, while cold dark matter (CDM) is unstable on all scales. We show for the first time that their relative drift triggers a resonant gravitational instability that drives sound waves in baryons. When the projected DM drift is subsonic, the stable oscillatory branch of baryons resonates with the Doppler-shifted DM mode, producing exponentially growing perturbations whose growth rates exceed the intrinsic CDM growth rate.
The instability peaks below the baryon Jeans scale and, in baryon-dominated environments, opens a window of complete stability between the Jeans scale and the resonance. Supersonic drift suppresses growth, as previously noted. The resonant coupling also transfers momentum between the species, creating a non-viscous, collisionless drag.
We derive an accurate analytical approximation for the growth rate at resonance and show that the associated timescales range from years to tens of millions of years across diverse environments---planets, protoplanetary disks, stars, molecular clouds, galaxies, and galaxy clusters---typically much shorter than their ages. In an expanding FLRW universe, the instability enhances baryon density perturbations at different redshifts for appropriately oriented modes while suppressing the growth of those aligned with the DM stream. The universe thus sings across all scales, and this resonant mechanism provides the means to listen: it offers a novel probe of dark matter through its seismic imprint on astrophysical objects and may explain long-standing puzzles such as the persistence of spiral arms and the heating of the intracluster medium in galaxy clusters.
\end{abstract}

\section{Introduction}

The growth of cosmic structure is driven by gravitational instability acting on small primordial density perturbations. In the standard $\Lambda$CDM paradigm, dark matter (DM) provides the dominant gravitational potential, while baryons follow under the influence of gravity and pressure forces. The classical Jeans instability \citep{Jeans1902,Binney+Tremaine_08} dictates that baryon perturbations are unstable only on scales larger than the Jeans length $\lambda_J = 2\pi c_s/\Omega_{\mathrm{b}}$, where $c_s$ is the sound speed, $\Omega_{\mathrm{b}}^2 =  4\pi G\rho_{\mathrm{b},0}$, and $\rho_{\mathrm{b},0}$ is the background baryon mass density. 
On smaller scales, pressure stabilizes baryons, and their perturbations propagate as sound waves. 
Cold DM, with negligible pressure, is unstable on all scales, growing at a rate $\sqrt{4\pi G\rho_{\mathrm{DM},0}}$ in the absence of baryons, where  $\rho_{\mathrm{DM},0}$ is the background DM mass density.
When baryons and DM coexist and are at rest relative to each other, the coupled linear system yields growth rates that interpolate between these limits: long-wavelength modes grow at $\sqrt{4\pi G(\rho_{\mathrm{b},0}+\rho_{\mathrm{DM},0})}$, while short-wavelength modes are stabilized by baryon pressure; however, a residual DM instability remains with a growth rate $\sim\sqrt{4\pi G\rho_{\mathrm{DM},0}}$ \citep[e.g.,][]{Dodelson2003,Weinberg2008}.

A crucial piece of physics that was long overlooked is the existence of a relative velocity between baryons and DM after recombination. Prior to decoupling, photons, baryons, and DM were tightly coupled via Thomson scattering and gravitational interactions. After recombination ($z\sim 1100$), photons decouple from baryons, and the baryon sound speed drops dramatically, while DM continues to stream freely. The result is a relative velocity that persists to low redshifts, with an rms value of about $30\,\mathrm{km\,s^{-1}}$ at recombination, decaying as $1/a$ thereafter~\citep{Tseliakhovich+Hirata_2010,Shimabukuro2023}. This streaming is a generic prediction of $\Lambda$CDM and has profound implications for structure formation on scales from the first stars to the large-scale distribution of galaxies~\citep{Binney+1998}.

The seminal work of \cite{Tseliakhovich+Hirata_2010} first demonstrated that this relative drift suppresses the growth of perturbations on scales comparable to the Jeans length since the DM speed is supersonic ($v_{\mathrm{DM}}  > c_s$). This suppression delays the formation of the first stars and affects the 21~cm signal \citep{Yoo2011,Fialkov2012, OLeary2012,Yacine+2014}. Subsequent studies have incorporated streaming velocities into cosmological simulations and shown that they impact galaxy formation, star formation histories, and the abundance of dwarf galaxies \citep[e.g.,][]{Bovy2013,Schmidt2016,Schauer+2019}. 
However, nearly all of this work has focused on the supersonic regime, where the drift suppresses the growth of baryon perturbations.

\begin{figure}
\includegraphics[width=1\linewidth]{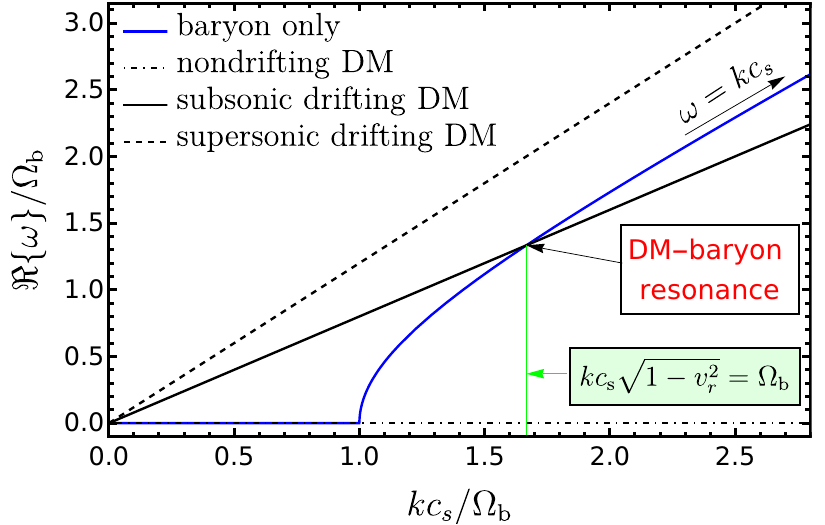}
\caption{\label{fig:Res}%
Graphical representation for the physical origin of the driven resonant instability. This figure shows the solutions of Equation~\eqref{Eq:disp00} in various limiting cases. The oscillation frequency for the baryon-only case ($\rho_{{\rm DM},0}=0$) is shown in blue, while the cold DM-only case is shown in black. For the non-drifting cold DM case (relative to the baryon rest frame), the normal modes are non-oscillating ($\omega=0$), represented by the dotted-dashed black line. When relative drift is included, the oscillation frequency is Doppler-shifted. The supersonic drift case ($v_r > 1$) is shown as a dashed black line, while the subsonic relative drift case is shown as a solid black line. This figure demonstrates that in the presence of both baryons and cold DM, a subsonic relative drift leads to resonance: at a particular wavelength, the DM and baryon oscillations become comparable.
We show in Section~\ref{sec:resonance} that this resonance drives sound waves in the baryons at wavelengths comparable to the resonance scale, with the growth rate peaking at this scale.
The resonant wave mode is such that $k c_s  \sqrt{1-v_r^2} = \OB$.
}
\end{figure}

The subsonic regime ($v_{\mathrm{DM}}\cos\theta < c_s$) has received comparatively little attention, despite being equally generic. For any given DM stream, there exist wave modes whose projected DM streaming speed is subsonic, especially those nearly perpendicular to the flow.
Moreover, the existence of a Doppler-shifted DM mode owing to the relative drift that can thus resonantly couple to baryon sound waves has, to our knowledge, never been explored (see a graphical illustration for the resonance in Figure~\ref{fig:Res}).
In this paper, we fill this gap by presenting a new resonant gravitational instability that operates precisely in the subsonic regime.

We show that when the projected relative drift is subsonic, the stable oscillatory baryon mode can resonate with the Doppler‑shifted cold DM mode at a specific wave mode $k_{\rm res}$ where their oscillation frequencies $\omega$ become equal. This resonance drives exponentially growing perturbations both at and close to $k_{\rm res}$, with the fastest growth rate exceeding the intrinsic DM growth rate. Remarkably, for baryon-dominated environments, this resonance opens a window of complete stability between the Jeans scale and the resonant scale, where both DM and baryon perturbations are stable—a feature absent in the standard picture. When the projected drift is supersonic, we recover the known suppression, thereby unifying both regimes within a single dispersion relation.

The instability we uncover is not merely a theoretical curiosity. The associated growth timescales range from years to tens of millions of years across diverse environments: planets, stars, protoplanetary disks, molecular clouds, galaxies, and galaxy clusters. In many cases, these timescales are much shorter than the ages of the objects, implying that the instability will be actively driving sound waves in these systems. In cosmology, incorporating this effect into linear perturbation theory for an expanding universe leads to enhanced baryon density perturbations at high redshifts when the projected drift is subsonic and suppression when it is supersonic—consistent with earlier results.

The resonant mechanism thus offers a new lens through which to view the interaction of dark and visible matter.
It opens a novel observational window: the driven sound waves may be detectable as characteristic seismic signatures in planets, stars, and ice sheets, offering a new route for dark matter detection.
Moreover, it could provide a natural explanation for long-standing puzzles such as the persistence of spiral structure in galaxies \citep{Toomre1964}
and may contribute to the heating of the intracluster medium (ICM), potentially ameliorating the cooling flow problem in galaxy clusters \citep{Fabian1994,McNamara+2007}.
A direct consequence of this resonant driving is a momentum exchange between the two species, which acts as a collisionless drag that can alter the bulk motion and energy balance in any environment where the instability operates.

The paper is organized as follows. Section~\ref{sec:disp} discusses the linear dispersion relation (derivations are presented in Appendix~\ref{app:disper}). Section~\ref{sec:resonance} analyzes the instability, focusing on cold DM and providing an analytical approximation for the growth rate (Appendices~\ref{app:nodrift} and~\ref{app:wdrift} discuss the cases of hot DM, without and with relative drift, respectively). Section~\ref{sec:tscale} evaluates the growth timescales across a wide range of astrophysical and cosmological environments—from planets and stars to molecular clouds, galaxies, and the expanding universe—showing that the instability can operate across vast scales, often with timescales much shorter than the ages of the objects. The section also addresses the role of DM coherence and its effect on the growth of perturbations. Section~\ref{sec:obervables} discusses observational prospects, including seismic signatures and cosmological probes. 
Section~\ref{sec:sandc} summarizes our findings and presents our conclusions.

\section{Dispersion Relation}
\label{sec:disp}

\begin{figure*}
\includegraphics[width=1.01\linewidth]{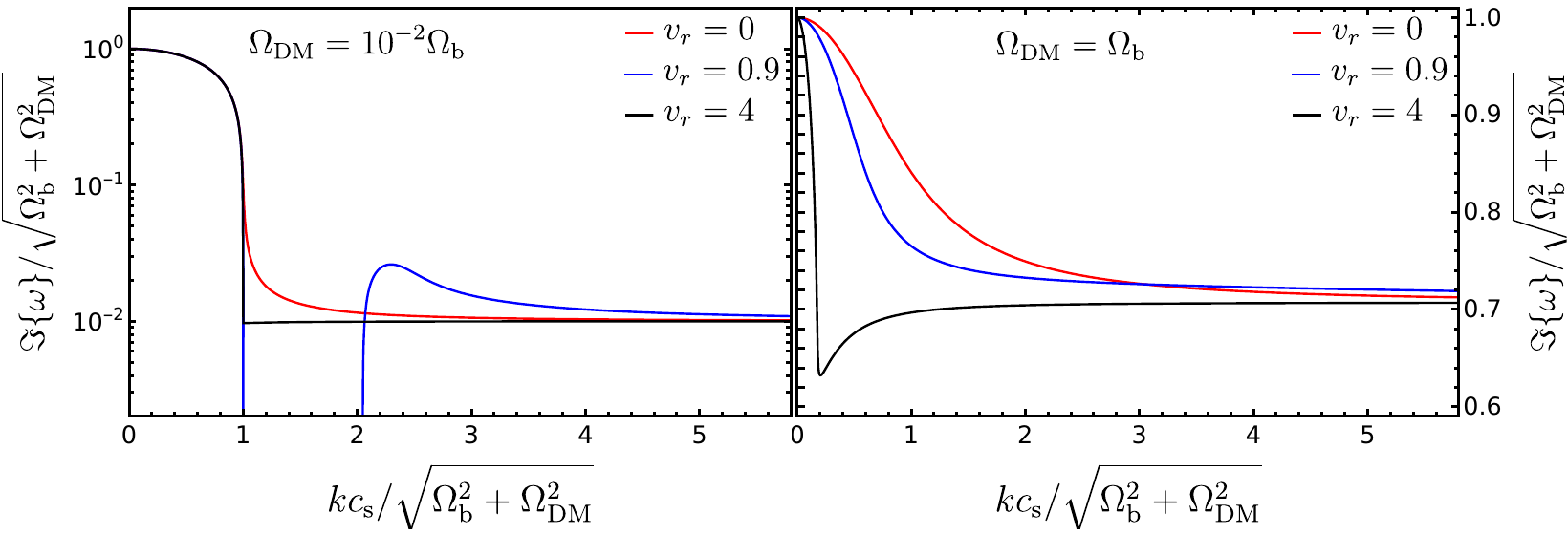}
\caption{\label{fig:Rvdr}%
The impact of dark matter (DM) drift on the gravitational field perturbation growth rates for two mass density ratios: $\Omega_{\mathrm{DM}} = 10^{-2}\Omega_{\mathrm{b}}$ (baryon-dominated, left panel, with logarithmic y-axes) and $\Omega_{\mathrm{DM}} = \Omega_{\mathrm{b}}$ (comparable densities, right panel, with linear y-axes). For each density ratio, growth rates are shown as a function of wavenumber for different relative drift velocities: subsonic drift $v_r = 0.9$ (blue), supersonic drift $v_r = 4$ (black), and the no-drift case (red) for comparison. In the baryon-dominated case (left), a subsonic projected drift creates a window of complete stability between the Jeans scale $k c_s/\Omega_{\mathrm{b}} = 1$ and close to the resonant scale ($k c_s/\OB \approx 2.2$), where the growth rate is exactly zero. At the resonant scale, the growth rate exceeds that of the no‑drift case. In the comparable‑density case (right), a subsonic drift (blue) also stabilizes long wavelengths relative to the no‑drift case (red), but it \emph{enhances} the growth rate at short wavelengths, with the blue curve rising above both the red and black curves for $k c_s > 3\sqrt{\Omega_{\mathrm{b}}^2 + \Omega_{\mathrm{DM}}^2}$. Supersonic drift (black curves) suppresses growth  consistent with previous findings \citep{Tseliakhovich+Hirata_2010}.
}
\end{figure*}

To derive the stability of linear perturbations in the baryon-DM coupled systems, we make the following assumptions.
Baryons are modeled as an ideal fluid with a uniform, stationary equilibrium density $\rho_{\mathrm{b},0}$ and pressure $p_{\mathrm{b},0}$. The small perturbations in baryons are adiabatic propagating with a constant sound speed $c_s$.
Dark matter is treated as a collisionless species described by a phase‑space distribution function; which in the cold limit for DM it reduces to a pressureless fluid.
The equilibrium configuration is taken to be homogeneous, static in the baryon rest frame, with the only bulk motion being a uniform relative drift $\vec{v}_{\mathrm{DM}}$ of the DM with respect to the baryons. Gravitational forces are governed by the Poisson equation, and we adopt the standard Jeans swindle: the zeroth‑order gravitational potential is set to zero even though the equilibrium densities are non‑zero. This ensures a consistent linearization around a homogeneous background.

Linear perturbations are taken to be plane waves. Solving the linearized fluid and Vlasov equations together with the perturbed Poisson equation yields a single dispersion relation that determines the frequency $\omega$ as a function of wavenumber $k$, the densities, sound speeds, and the relative drift. The value of $\omega$ dictates the behavior of the gravitational potential and, consequently, the evolution of both DM and baryon density perturbations. For a given solution $\omega$, its real part gives the oscillatory frequency of the perturbations at wavenumber $k$, while its imaginary part determines the exponential growth or damping rate, depending on the sign.

The full derivation for the dispersion relation is presented in Appendix~\ref{app:disper}.
It is given by
\begin{eqnarray}
-1 
&=& 
\frac{\Omega_\mathrm{b}^2 }
{ \omega^2 - k^2 c_s^2}+ 
\frac{\Omega_\mathrm{DM}^2 }{
(\omega - \vec{k} \cdot \vec{v}_\mathrm{DM})^2 - k^2 c_\mathrm{DM}^2},
\label{Eq:disp00}
\end{eqnarray}
where the baryon and DM gravitational frequencies are $\Omega_\mathrm{b}^2 \equiv 4 \pi G  \rho_{b,0}  $ 
and $\Omega_\mathrm{DM}^2 \equiv 4 \pi G  \rho_{\mathrm{DM},0}$, respectively; $c_s$ and $c_\mathrm{DM}$ are the sound speed in the baryons and the velocity dispersion of the DM, respectively.
We note here that the solution of the dispersion relation gives the oscillatory or growing behavior of the first order gravitational perturbation $\phi_1$, which is also the behavior of the initial mass density perturbation in various species normalized to their initial value, i.e., $\rho_{1,\mathrm{b(DM)}}/\rho_{0,\mathrm{b(DM)}} \propto \phi_1 $.

To investigate the solutions for this dispersion relation, We define the following dimensionless variables:
$x = \omega/\Omega_\mathrm{b}$, $y = k c_s / \Omega_\mathrm{b}$, $R = \Omega_\mathrm{DM}/\Omega_\mathrm{b}$, $c_r = c_\mathrm{DM}/c_s$, which is the ratio of the DM sound speed to the baryon sound speed, and $v_r = v_\mathrm{DM} \cos(\theta)/c_s$, which is the projected DM propagation speed normalized by the baryon sound speed $c_s$, and
$\theta$ is the angle between the DM propagation direction and the sound wave propagation direction ($\vec{k}$).
Thus, in these normalized quantities, the dispersion relation simplifies to
\begin{eqnarray}
-1 = 
\frac{1}{ x^2 - y^2}+ 
\frac{R^2}{
(x - y v_r)^2 - y^2 c_r^2}.
\label{Eq:disp01}
\end{eqnarray}

\section{Impact of relative drift}
\label{sec:resonance}

In this section, we consider the cases of $v_r \neq 0$ in the cold DM case, i.e, $c_r=0$.
In the Appendix~\ref{app:nodrift}, we discuss the solutions of the dispersion relation in the cases of $v_r=0$.
In the case of long wavelength, i.e., $y \rightarrow 0$, the value of $v_r$ does not affect the solutions and yields the same result as in the case of $v_r=0$ discussed in the Appendix~\ref{app:nodrift}, namely $\Im\{\omega\} = \sqrt{\ODM^2+\OB^2}$.
We discuss here the case for cold DM, i.e., $c_r=0$.
The impact of such drift in the hot DM case on the resonant instability is presented in Appendix~\ref{app:hdrft}.

We begin by showing the solutions of the dispersion relation in Equation~\eqref{Eq:disp00} for various limiting cases in Figure~\ref{fig:Res}. The oscillation frequency for the baryon-only case ($\rho_{\mathrm{DM},0} = 0$) is shown as a blue line. This represents the expected behavior of baryons: very long wavelength modes are unstable and non‑oscillatory, while at short wavelengths, the perturbations behave as stable sound waves with $\omega \sim k c_s$.
For cold DM, all wave modes are unstable and non‑oscillatory; we show these cases with black lines. In the rest frame of cold DM, the solutions exhibit non‑oscillating modes ($\omega = 0$, dot‑dashed black line). A relative drift speed with respect to baryons Doppler‑shifts the DM oscillation frequency. An example for the case of supersonic drift ($v_r > 1$) is shown as a dashed black line, while an example for a case with subsonic drift is shown as a solid black line.

The physics of the driven resonant instability is illustrated in Figure~\ref{fig:Res}: when both baryons and cold DM are present, a sonic or subsonic relative drift causes the DM (solid black line) and baryons (blue line) to oscillate with the same frequency at a particular wavelength, producing a resonance. This resonance drives sound waves in the baryons at the corresponding scale, with the instability growth rate peaking at the resonant wavelength, as shown below.

\textit{We emphasize here that $v_r \equiv v_{\mathrm{DM}} \cos\theta / c_s$ is a measure of the projected relative drift compared to the baryon sound speed.}
The resonant wave mode can be simply derived by noting that the baryon-only solution ~\citep{Binney+Tremaine_08} is given by $\omega^2 =   k^2 c_s^2 -  \OB^2$, while the Doppler-shifted DM oscillation frequency is $w = k v_{\rm DM} \cos \theta$. We can then find the wave mode at which these two oscillation frequencies are equal:
$ k^2 c_s^2 - \OB^2 = (k v_{\rm DM} \cos \theta)^2 $, which gives the resonant wave mode such that $k c_s  \sqrt{1-v_r^2} = \OB$.

In Figure~\ref{fig:Rvdr}, we present the growth rate as a function of wavenumber $k$ (obtained by solving the dispersion relation in Equation~\eqref{Eq:disp00}) for various values of $R = \ODM/\OB$ and $v_\mathrm{r}$.
Two distinct regimes emerge, separated by the value of $R$:

Firstly, in the baryon-dominated case (left panel of Figure~\ref{fig:Rvdr}), when the projected dark matter velocity in the direction of the sound wave is subsonic, wave modes between the Jeans scale $k c_s/\Omega_{\mathrm{b}} = 1$ and close to the resonant scale (here $k c_s/\Omega_{\mathrm{b}} \approx 2.29$) exhibit complete stability in both baryon and dark matter perturbations, as the growth rate of gravitational perturbations is exactly zero in this window. At the resonant scale, the instability growth rate exceeds that of the no-drift case (red curve). This effect vanishes entirely when the relative drift becomes supersonic (black curves).

Secondly, when the cold dark matter density is comparable to the baryon density (right panel of Figure~\ref{fig:Rvdr}), both subsonic and supersonic drifts enhance stability for wavelengths comparable to or longer than the Jeans scale, as seen by the blue and black curves lying below the red (no-drift) curve; they exhibit slower growth rates. 
Moreover, in this regime, a subsonic relative drift enhances the growth rate of short wave modes, as indicated by the blue curve rising above both the red and black curves for $k c_s > 3\sqrt{\OB^2 + \ODM^2}$.
The supersonic drift leads to the suppression of the growth rate, especially for wavelengths longer than the Jeans scale ($k < k_J$), as evidenced by the black curve being well below the red curve.
This is consistent with the previous prediction by~\citet{Tseliakhovich+Hirata_2010}, who studied this case in an expanding universe.

\section{Impacts of the resonance on various astrophysical to cosmological scales}
\label{sec:tscale}

In the previous sections, we have shown that a relative drift between DM and baryons can significantly modify the traditional Jeans instability.  Two key features that will inform the astrophysical implications of this drift are: first, the fastest growth of the resonant wave modes is faster than the corresponding cold DM growth rates in the absence of drift, and second, wave modes lying between the resonance scale and the baryon Jeans scale can be stabilized in the baryon dominated cases ($R < 1$).
We now turn to estimating specific instability growth times, resonance scales, and their dependence on the astrophysical environment before discussing their application to a number of astronomical environments.

\subsection{Resonance time and length scales}
\label{sec:Tres}

\begin{figure}
\includegraphics[width=1.02\linewidth]{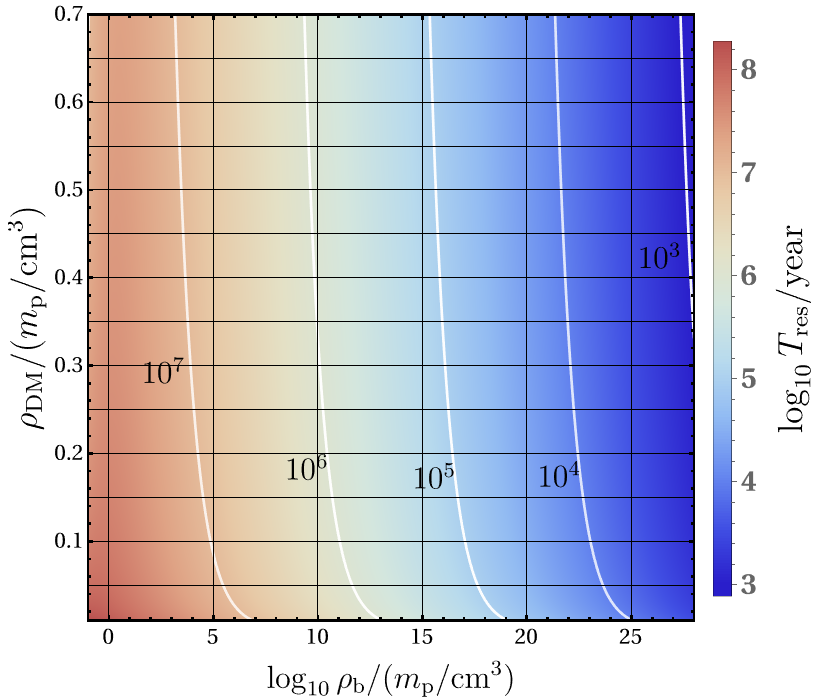}
\caption{\label{fig:TRes}%
The typical time scale for the resonantly driven sound waves as given in Equation~\eqref{eq:Tres}, where we use $v_r=0.5$.
We normalize the both cold DM baryon mass densities with the Interstellar medium average density of 1 proton (with mass $m_\mathrm{p}$) in ${\rm cm}^3$.
The upstreamgrowth time can be much shorter if the value of $v_r$ is chosen differently; see discussion at the end of section~\ref{sec:Tres}.
White contours show the lines of equal growth times in years.
Here, $\rho_{\mathrm{b}}$ and $\rho_{\mathrm{DM}}$ refer, respectively, to the baryon and DM background densities.
}
\end{figure}

We start by computing the physical time for the exponential growth of the driven sound waves.
For the case of cold DM, the growth rate at the resonance scale 
\begin{eqnarray}
k_\mathrm{res} c_s \sqrt{1-v_r^2} = \OB
\quad \Rightarrow \quad
k_\mathrm{res} c_s = \OB \sqrt{1+ u_r^2},
\qquad \qquad
\label{eq:kres}
\end{eqnarray}
is given by (see Appendix~\ref{app:cdrft})
\begin{eqnarray}
   \Gamma_{\rm res} = \frac{ \left(R^{8/5}+1\right)}{\sqrt[6]{9 u_r^2+R}} R^{2/3} \OB
     =
     \frac{ \left(R^{8/5}+1\right)}{\sqrt[6]{9 u_r^2+R}} \OB^{1/3} \ODM^{2/3},
     \qquad
     \label{Eq:grate}
\end{eqnarray}
where $u_r = v_r/\sqrt{1-v_ r^2}$.
This is a good approximation only when $R \leq 1$ ($\ODM  \leq \OB$).
While this formula is inspired by the asymptotic behavior of the growth rate when $R \ll 1$, the above formula provides an accurate estimate for all values of $R \leq 1$. The only caveat of this point is that this growth rate $\Gamma_{\rm res}$ is accurate only when $\Gamma_{\rm res} > R ~\OB$. We discuss this thoroughly in Appendix~\ref{app:cdrft}.
The growth time for the resonantly driven sound waves in such a case is given by

\begin{eqnarray}
T_{\rm res} &=& \frac{1}{ \Gamma_{\rm res} }
=
\frac{\sqrt[6]{9 u_r^2+R}}{ \left(R^{8/5}+1\right)}
\OB^{-1/3} \ODM^{-2/3}
\nonumber \\
&=&
\frac{\sqrt[6]{9 u_r^2+R}}{  \sqrt{4 \pi G}  \left(R^{8/5}+1\right)}
\rho_{\mathrm{b},0}^{-1/6} \rho_{\mathrm{DM},0}^{-1/3}
\nonumber \\
&\approx&
3783 
\frac{\sqrt[6]{9 u_r^2+R}}{ \left(R^{8/5}+1\right)}
\left( \frac{ \rho_{\mathrm{b},0} }{{\rm g}/{\rm cm}^3} \right)^{-\tfrac{1}{6}} 
\left( \frac{ \rho_{\mathrm{DM},0}}{ \rho_{\mathrm{DM},*} } \right)^{-\tfrac{2}{6}}
~ {\rm yr}
\nonumber \\
&\approx&
3.47\times10^7\,
\frac{\sqrt[6]{9 u_r^2+R}}{ \left(R^{8/5}+1\right)}
\left( \frac{ \rho_{\mathrm{b},0} }{m_\mathrm{p}/{\rm cm}^3} \right)^{-\tfrac{1}{6}} 
 \left( \frac{ \rho_{\mathrm{DM},0}}{ \rho_\mathrm{DM,*} } \right)^{-\tfrac{2}{6}}
~ {\rm yr}
\nonumber \\
\label{eq:Tres}
\end{eqnarray}
Where,
$R = \sqrt{ \rho_{\mathrm{DM},0}/ \rho_{\mathrm{b} ,0}}$, 
$m_\mathrm{p} = 1.6 \times 10^{-24} $ g is the proton/ion mass, and $\rho_{\mathrm{DM},*} = 0.43~ {\rm GeV}/{\rm cm}^3 c^2 = 7.6 \times 10^{-25} {\rm g}/{\rm cm}^3$ is the average mass density for DM in the solar neighborhood; that is, $\rho_{\mathrm{DM},0}/\rho_{\mathrm{DM,}*}$ represents the local fractional increase or decrease in the mass density of DM compared to the average mass density in the solar neighborhood.
These relevant time scales are visualized in Figure~\ref{fig:TRes}.
In such a case,
the most unstable wavelength is 
\begin{eqnarray}
    \lambda_{\rm res} = \frac{2 \pi c_s \sqrt{1- v_r^2}}{\OB} = \lambda_J \sqrt{1-v_r^2} 
    = \frac{\lambda_J}{ \sqrt{1+u_r^2}},
    \label{eq:resL}
\end{eqnarray}
where, $v_r \equiv v_{\rm DM} \cos \theta /c_s <1$ is the projected relative drift divided by the baryon's sound speed $c_s$, and 
$\lambda_J = 2 \pi c_s/\OB$ is the baryon's Jean's wavelength.
When the relative drift is very subsonic, i.e., $v_r \ll 1$, the resonant wavelength is very close to the Jeans wavelength in the baryons.
However, when the drift close to sonic drift, i.e., $u_r \gg 1$, the resonant wavelength can be very small.
This means that the wavelength for the fastest growing sound waves depends on the sound  speed in the baryons directly through $\lambda_J$.
The oscillation frequency for that most unstable wave depends on $c_s$ only indirectly through $v_r$:
\begin{eqnarray}
\label{eq:wres}
    \omega_{\rm res} = 
    \sqrt{ k_{\rm res}^2 c_s^2 - \OB^2 }
    =
    \sqrt{ \OB^2 (1+u_r^2) - \OB^2}
    =
    u_r \OB \qquad
\end{eqnarray}

For relative drift very close to the sound speed, i.e., $u_r \gg 1$, the oscillation frequency becomes much larger than $\Omega_{\mathrm{b}}$ because resonance then occurs at a wavenumber $k$ that is significantly larger than the Jeans wavenumber $k_J = \Omega_{\mathrm{b}}/c_s$, and thus $ \lambda_{\rm res} \ll \lambda_J$.

\subsection{Important Characteristics of the Instability}
\label{sec:charac}

This new instability thus has a number of important characteristics that we highlight here:

\begin{figure}
\includegraphics[width=1.0\linewidth]{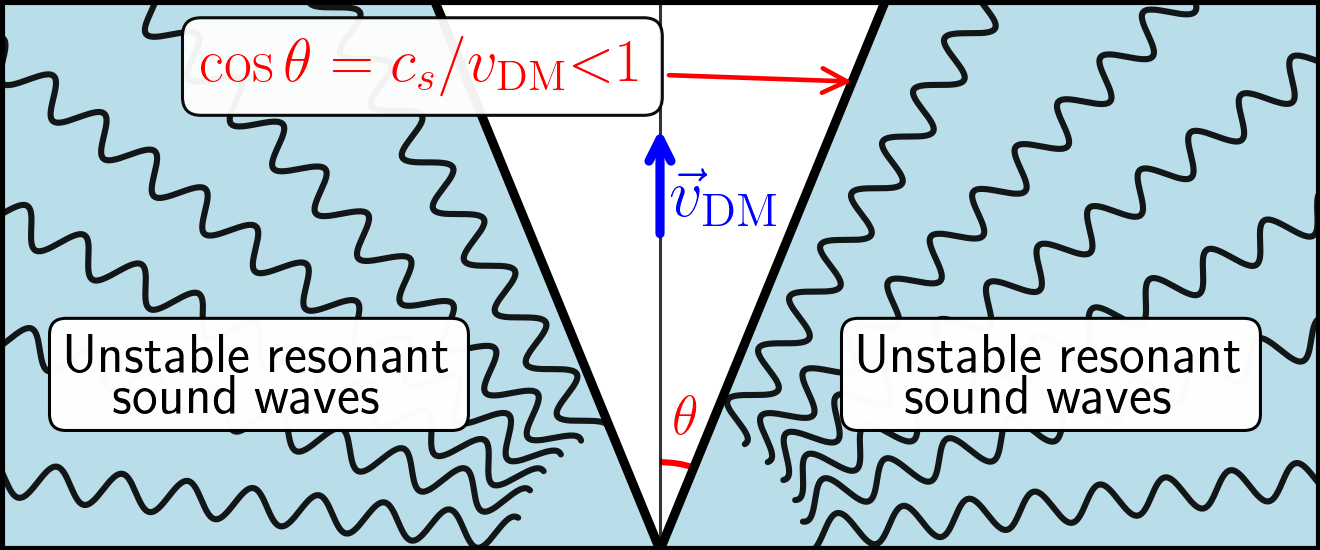}
\caption{\label{fig:costheta}%
Regime of unstable sound waves (light blue shaded regime) due to resonant instabilities in case of supersonic DM relative drift, i.e., $v_\mathrm{DM} > c_s$ along the blue vertical direction. In case of subsonic drift, the cone of stable sound waves closes, and sound waves are resonantly unstable in all direction.
Modes are stable on scales below the Jeans scale of Baryons.
}
\end{figure}

\begin{itemize}

\item The resonance exists only when \(0 < v_r < 1\); that is, the dark matter must have a non-zero relative drift with respect to the baryons, and the projected drift on the wave mode must be subsonic (\(v_r \equiv v_{\mathrm{DM}} \cos\theta / c_s < 1\)). 
Consequently, this resonance occurs only at wavelengths shorter than the Jeans length of the baryons.
When \(v_r \ll 1\), the resonance appears on scales close to the Jeans scale, whereas when \(v_r \sim 1\), it occurs at wavelengths much smaller than the Jeans scale.

\item When \(v_{\mathrm{DM}} < c_s\), sound waves in all directions relative to the dark matter propagation direction are resonantly unstable.
In the supersonic case (\(v_{\mathrm{DM}} > c_s\)), only modes whose projected drift is subsonic are subject to the resonant instability; these are modes with an angle \(\theta\) satisfying \(\cos \theta \leq c_s / v_{\mathrm{DM}}\).
Figure~\ref{fig:costheta} provides an illustration of this projection.

\item A single drifting dark matter stream destabilizes multiple wave modes, not merely a single mode. 
This also implies that when multiple DM streams coexist spatially, different streams with various propagation directions can excite the same wave mode simultaneously, provided that each stream satisfies \(v_r = v_{\mathrm{DM}} \cos\theta / c_s < 1\) (where each stream may have different values of \(v_{\mathrm{DM}}\) and \(\cos\theta\)).
    
\item Our analytical approximation for the growth rate in Equation~\eqref{Eq:grate} provides important intuition about the instability across various regimes—intuition that is difficult to discern directly from the dispersion relation. For fixed values of $R$ and $v_{\mathrm{DM}}/c_s$, the quantity $v_r$ varies with the angle of the wave mode relative to the DM streaming direction, ranging from $0$ to $v_{\mathrm{DM}}/c_s$ (or to $1$ if $v_{\mathrm{DM}}/c_s > 1$). Consequently, one can identify the angle that maximizes the growth rate, i.e., that minimizes the growth time. For a given $R$, the growth rate is maximized when $9u_r^2 < R$, which is equivalent to $u_r < \sqrt{R/9}$. This behavior is evident in Appendix~\ref{app:wdrift}: once $u_r < \sqrt{R/9}$, the growth rate attains its maximum and becomes independent of $u_r$. For $u_r > \sqrt{R/9}$, shorter‑wavelength sound waves (see Equation~\ref{eq:resL}) are also driven, albeit at lower growth rates, i.e., on a longer time scale.

\item The instability described here is the gravitational analog of the well known cold beam–plasma instability. The similarities include: when DM propagates at a speed much larger than the sound speed, the fastest-growing modes are oblique, analogous to the relativistic beam–plasma case \citep[e.g.,][]{bret-2010-pop,linear-paper,resolution-paper}, which is exactly the behavior discussed in the point above. Moreover, the normalized fastest growth rate in beam–plasma instabilities is $\Gamma \propto (n_b/n_g)^{1/3}$, where $n_b$ and $n_g$ are the beam and background number densities, respectively \citep{bret-fluid-2006,blazari}. In our case, the normalized growth rate is $\Gamma \propto R^{2/3} = (\rho_{\mathrm{DM},0}/\rho_{\mathrm{b},0})^{1/3}$. 
In both cases, the growth rate is measured in units of the fundamental frequency, i.e., plasma frequency in the case of beam-plasma instabilities, and the gravitational frequency in the instability discussed in this work.

\item The growth time at the resonance scale depends weakly on the baryon and DM mass densities: $T_{\rm res} \propto \rho_{\mathrm{b},0}^{-1/6} \rho_{\mathrm{DM},0}^{-1/3}.$

\end{itemize}

In what follows, for compactness we drop the subscript $_0$ in both baryon and DM background densities.  That is, henceforth, $\rho_{\mathrm{b}}$ and $\rho_{\mathrm{DM}}$ refer to the baryon and DM background densities, respectively.

\subsection{Impacts of the resonant instability}
\label{sec:impacts}

\begin{table*}[!htbp]
\centering
\caption{%
Typical parameters for different astrophysical environments. Baryon mass density \(\rho_{\mathrm{b}}\) is normalized to the typical ISM density of a proton (mass \(m_{\mathrm{p}}\)) per \(\mathrm{cm}^3\); DM mass density is normalized to \(\rho_\mathrm{G} = \mathrm{GeV}/(\mathrm{cm}^3\,c^2)\). The gravitational frequencies ratio is \(R = \Omega_{\mathrm{DM}}/\Omega_{\mathrm{b}} = \sqrt{\rho_{\mathrm{DM}}/\rho_{\mathrm{b}}}\).
Again, $\rho_{\mathrm{b}}$ and $\rho_{\mathrm{DM}}$ refer, respectively, to the baryon and DM background mass densities.
We choose the value of \(u_r\) that maximizes the growth rate while respecting the finite extent of each object, i.e., using Equation~\eqref{eq:ur}. Growth times \(T_{\mathrm{res}}\) and frequencies \(\omega_{\mathrm{res}}\) are computed by directly solving the dispersion relation at the resonance scale (Equation~\ref{Eq:disp02}). The wavelength of the driven sound waves is \(\lambda_{\mathrm{res}}\); for all cases, our choice of \(u_r\) ensures \(\lambda_{\mathrm{res}} \leq L\), where \(L\) is the typical size adopted for each environment. The quantity \(\omega_{\mathrm{res}}/\Gamma_{\mathrm{res}}\) is the number of oscillations during one \(e\)-fold growth.
Using Equation~\eqref{eq:prop}, when \(\omega_{\mathrm{res}}/\Gamma_{\mathrm{res}} \gg 1\), the driven (unstable) waves propagate beyond the typical extent of the object; when \(\omega_{\mathrm{res}}/\Gamma_{\mathrm{res}} < 1\), the waves are confined within the environment.
Circumstances that might suppress or prevent the growth of the driven waves are discussed in Section~\ref{sec:applicability}.
All growth times, resonant wavelengths, and frequencies listed in this table are examples for the possible modes and are an order-of-magnitude estimates obtained by applying the homogeneous dispersion relation (Equation 1) with local parameters appropriate to each environment.
}
\label{tab:param}
\begin{tabular*}
{\textwidth}{@{\extracolsep{\fill}}
lcccccccccc
@{}}
\toprule
Environment & 
$\dfrac{ \rho_{\rm DM} }{\rho_\mathrm{G}}$& 
$\dfrac{ \rho_{\rm b}  }{m_\mathrm{p}/{\rm cm}^3}$
& $ R $
& $\dfrac{ c_s }{{\rm km\,/\,sec}}$
& $ u_r $  
& $T_{\rm res}$ (yr) 
&
$\omega_{\rm res}$ (Hz) 
&
$\omega_\mathrm{res} /\Gamma_\mathrm{res}$
& $\lambda_\mathrm{res}$ 
\\
\midrule
Planets   
& 0.43   
& $6 \times 10^{24} $  
& $2.7 \times 10^{-13} $ 
& $7$
& $0.97$  
& $3715$ 
& $0.0028$ 
& $ 3.3\times 10^{8} $
& $10886$ km
\\
Ice sheets &  $0.43$     &  
$6\times 10^{23}$   & 
$ 8.7 \times 10^{-13}$ 
& $3.5$
&  $18$ 
& $14.4\times 10^3$ 
& $0.016$ 
& $ 7.4 \times 10^{9} $ 
& 1336 km
\\
Protoplanetary disk      & 
0.43   & $6 \times 10^{11}$  & 
$ 8.7\times 10^{-7}$ 
& $0.5$ 
& $0.00046$
& $40.5  \times 10^3$
& $2.4 \times 10^{-13}$
& $ 0.3 $ 
& 22.9 Au
\\
Stars: sun  
& $0.43$     
& $8.3\times 10^{23}$ 
& $ 7.4 \times 10^{-13} $   
& $450$ 
& $2.36$ 
& $6928 $ 
& $0.0025$ 
& $ 5.6 \times 10^{8} $
& $ 10^6$ km
\\
Stars: white dwarfs  & 
0.43     &  $6\times 10^{29}$  
& $8.7 \times 10^{-16}$
& $3000$
& $1.6$
& $646$ 
& $1.48$
& $ 3 \times 10^{10}  $
& $10^4$ km
\\ 
Stars: neutron stars  
& 0.43    
&  $6\times 10^{38}$  
& $2.8\times 10^{-20}$ 
& $10^5$
& $0.63$
& $14.9$ 
& $18.3 \times 10^{3} $
& $ 8.6  \times 10^{12} $
& $18$ km
\\
GMCs: dense        & 
$0.43 $  & $10^6$ & 
$ 6.7\times 10^{-4}$ 
& $0.5$
& $0.013$
& $ 1.1 \times 10^6$ 
& $8.6 \times 10^{-15}$& 
$0.3$
& $0.086$ pc
\\
GMCs: light 
& $0.43$  
& $10^2$  
& $6.7 \times 10^{-2}$ 
& $0.2$
& $0.13$
& $ 1.1 \times 10^7$ 
& $8.9\times 10^{-16}$ 
& $0.3$
& $3.4$ pc
\\
Galaxies   
& $0.35$
& $1$ 
& $ 0.61 $ 
& $10$
& $0.39 $
& $31 \times 10^6$ 
& $3.2 \times 10^{-16}$ 
& $0.32$
& $1.6$ kpc
\\
Galaxy clusters  
& $ 0.0035$ 
& $ 0.0004$  
& $ 3.0 $ 
& $300$
& $3.5 $ 
& $ 433 \times 10^6 $ 
& $ 8.2 \times 10^{-17}$ 
& $1.13 $
& $700$ kpc
\\
\bottomrule
\end{tabular*}
\vspace{0.0cm}
\end{table*}

DM–baryon relative drifts are ubiquitous across astronomical environments, from the early universe to terrestrial scales. They originate at decoupling ($z\sim1000$) with a very large coherent scale of $\sim200\,\mathrm{Mpc}$ and are initially supersonic \citep{Tseliakhovich+Hirata_2010}. As the universe expands, DM collapses into halos, and galaxy clusters and galaxies form inside these DM halos \citep{Wechsler+2018}. Consequently, the relative drift persists within and around these structures, influencing their evolution.

Inside galaxies, the DM halo breaks into subhalos, and the DM becomes virialized, producing multiple streams at each location. This can be effectively modeled as a kinematically heated distribution, i.e., with a velocity dispersion. Although our main analysis focuses on the cold DM (CDM) case, we derive results for kinematically hot DM in Appendix~\ref{app:hdrft}, showing that the resonant growth persists. Intuitively, multiple streams can each excite multiple wave modes simultaneously. For any single stream, the instability may grow at a reduced rate compared to that of a single coherent stream, analogous to beam–plasma instabilities \citep{chang:2014}. However, because multiple streams coexist, the effective growth rate approaches the maximum growth rate achievable from a single stream.

One might naively assume an upper limit of \(v_r = v_{\mathrm{DM}}/c_s\), which is set by the local DM speed and the baryon sound speed. However, near collapsed objects such as stars, gravitational acceleration increases the DM infall speed, raising that limit and making the full range of resonant conditions accessible. More generally, the DM drift velocities relevant for the environments below are no longer the primordial drifts from the early universe; instead, they are set by the kinematics of virialised halos and subhalos (e.g., rotation, infall, or mergers).  
For pressure‑supported objects much smaller than the DM halo scale, such as stars, gravitational acceleration generically leads to supersonic infall velocities, opening the full range of resonant instability conditions  (typically both supersonic and subsonic drift are realized) independently of the background DM dynamics.
The Sun illustrates this: the solar neighborhood DM speed is \(220\) km/s, and the conservation of energy yields the speed at the Sun’s surface: $v_{\mathrm{DM},\odot}=v_\mathrm{DM}\sqrt{1+2GM_\odot/(R_\odot v_\mathrm{DM}^2)}=565$ km/s. While the neighborhood speed is below the average sound speed inside the Sun ($\sim450$ km/s), gravitational infall makes DM supersonic at the surface.

Our calculation assumes an infinite, uniform background density for both DM and baryons. For self-supported, self-gravitating objects (e.g., stars and galaxies), this assumption is not well justified. Nevertheless, the underlying physical picture, and therefore the qualitative impact, of our new instability remains unchanged for two reasons. First, the wavemode of such objects is constrained such that $\omega\approx c_s k$ at sufficiently high $k$, and necessarily approaches zero near the minimum wavenumber accessible to the system. As a result, the Doppler‑boosted DM dispersion relation always crosses the sonic branch and instability arises. Second, while the dispersion relation is usually discrete for systems with finite extent, density gradients broaden the instability resonance scale \citep[i.e., the width of the blue bump near $kc_s \approx 2.2\sqrt{\Omega_{\mathrm{b}}^2+\Omega_{\mathrm{DM}}^2}$ in the left panel of Figure~\ref{fig:Res}, see e.g.,][]{sim_inho_18, th_inho_20}.
This is typically sufficient to capture at least one mode.\footnote{The presence of density gradients separates this case from that presented in \citet{resolution-paper}. Uniform density objects with finite extent may not be unstable, but such objects are hard to imagine occurring naturally.} Quantifying the impact of these effects in various environments is left to future work.
Here, we make order-of-magnitude estimates estimates for the fastest growing wavelength, oscillation frequency, and growth timescale employing the homogeneous dispersion relation for a variety of astrophysical environments.

To estimate the time and length scales over which we expect the consequences of our new instability to manifest in self‑supported astronomical objects, we make the conservative choice to restrict the longest resonant wavelength to the object size $L$. If the Jeans scale $\lambda_J$ is larger than $L$, the longest available resonant wavelength is $\lambda_{\rm res}^{\rm max} = L$. This sets a minimum value of $u_r$,
\begin{eqnarray}
\label{eq:urmin}
u_{r,{\rm min}} 
=
\begin{cases}
\sqrt{ (\lambda_J/L)^2-1}, & \lambda_J \leq L,\\
0, & \lambda_J < L,
\end{cases}
\end{eqnarray}
which must be respected when choosing $u_r$ to maximize the growth rate.\footnote{As noted previously, due to gravitational infall, the DM generally allows $u_r\ge u_{r,{\rm min}}$.} Therefore we set

\begin{eqnarray}
\label{eq:ur}
u_r =  1.5 \max (u_{r,{\rm min}}, \sqrt{R/9}),
\end{eqnarray}
where the pre‑factor $1.5$ (somewhat arbitrary) ensures $\lambda_\mathrm{res} < L$. 
Despite this quantitative modification at the largest scales, the general physics of the instability – including the growth rates at scales shorter than the object – remains unchanged.

The distance that the resonant wave mode propagates during one e‑fold growth time, $l_\mathrm{pr}$, (normalized by $\lambda_{\mathrm{res}}$)  is 
\begin{equation}
\frac{l_\mathrm{pr}}{\lambda_{\mathrm{res}}}
=
\frac{T_{\mathrm{res}} v_{\mathrm{ph}}}{\lambda_{\mathrm{res}}}
= \frac{T_{\mathrm{res}} \omega_{\mathrm{res}}}{k_{\mathrm{res}} \lambda_{\mathrm{res}}}
= \frac{\omega_{\mathrm{res}}}{2\pi \Gamma_{\mathrm{res}}}.
\label{eq:prop}
\end{equation}
In Table~\ref{tab:param}, we show the values of $\omega_{\mathrm{res}} / \Gamma_{\mathrm{res}}$ for various environments. Since $\lambda_{\mathrm{res}} \le L$,
a value $\omega_{\mathrm{res}} / \Gamma_{\mathrm{res}} \gg 1$ indicates that the driven waves propagate beyond the object’s extent, while $\omega_{\mathrm{res}} / \Gamma_{\mathrm{res}} < 1$ implies that the waves are confined within the environment.

Typical parameters for various environments are summarized in Table\,\ref{tab:param},\,%
including growth times and oscillation frequencies computed directly from the dispersion relation (Equation~\eqref{Eq:disp00}), which agrees well with the approximation in Equation~\eqref{eq:Tres}.
For planets, protoplanetary disks, stars, and molecular clouds, we adopt the solar neighborhood DM density $\rho_{{\rm DM}}=0.43\,\mathrm{GeV}/c^2\,\mathrm{cm}^3$ \citep{McKee+2015,Gaia_rhoDM-2025}; for the ISM of the Milky Way, $\rho_{\rm DM}=0.35\,\mathrm{GeV}/c^2\,\mathrm{cm}^3$ \citep{Kafle+2014}; for galaxy clusters, both baryon and DM densities are taken to be $10^3$ times the cosmic mean at $z=0$ (with $H_0=71\ \mathrm{km\,s^{-1}\,Mpc^{-1}}$, $\Omega_{b,0}=0.044$, $\Omega_{\mathrm{cdm},0}=0.226$). 
The following subsections expound on the potential impacts of the instability in these environments.

\subsubsection{Planets}
\label{sec:planets}

For planets, we take a typical mass density of\,$10~\mathrm{g\,cm^{-3}}$\, and an object size of $L=1.27\times 10^3$ km
(Earth diameter), which is appropriate for Earth-like planets. The growth time for the fastest driven vibrational mode is about $3700$ years, much shorter than the nominal 4.5 billion-year age of the solar system and the geological timescale of Earth.

The oscillation frequency of the most unstable wave mode is 2.8 mHz, with a wavelength of $\sim 10^4$ km.
Higher frequency (shorter wavelength) wave modes are also driven on a longer timescale by our new instability.

We discuss further in Sections~\ref{sec:ice} and~\ref{sec:seismology} the potential observable seismic signatures and/or vibrational oscillations, both in the Earth itself and on continental glaciers.

\subsubsection{Protoplanetary disks}

Protoplanetary disks of gas and dust orbit young stars, providing the raw material for planet formation~\citep{williams2011protoplanetary}. They persist for 3–10 Myr with initial masses of 0.03–0.3 M$_\odot$~\citep{manara2023protoplanetary}. Although gas dominates, dust ($\sim$1\% of the mass) supplies the solids for planet assembly, and ALMA observations reveal substructures (rings, gaps, asymmetries) tracing ongoing planet formation~\citep{andrews2020observations}.
Within these disks, planet formation proceeds in stages: micrometer‑sized dust grains settle to the midplane and coagulate into larger aggregates~\citep{testi2014dust}; streaming instability and pebble accretion then form planetesimals and planetary cores~\citep{johansen2015planetesimals}. When a core reaches $\sim 10\,\text{M}_{\odot}$, it rapidly accretes gas to become a giant planet~\citep{pollack1996formation}.

The typical surface density for these disks is $\approx 200\,\mathrm{g\,cm^{-2}}$, with a scale height of $15\,\mathrm{AU}$ and a radius of $250\,\mathrm{AU}$, yielding a typical volume density of $\rho = 10^{-12}\,\mathrm{g\,cm^{-3}} \approx 5.9 \times 10^{11} m_\mathrm{p}\,\mathrm{cm^{-3}}$ \citep{pollack1996formation}.

We restrict ourselves to the finite extent of $L = 250\,\mathrm{AU}$ and predict that a relative drift of order $5\times 10^{-4} c_s$ will excite sound waves on timescales of approximately $40\,\mathrm{kyr}$ with a wavelength of about $23\,\mathrm{AU}$. Since this is much shorter than typical ice sheet lifetimes, the new instability excites such density waves and also shorter wavelength modes on longer timescales.
This can potentially enhance mixing during planetary core formation and accelerate the growth of pebbles and planetesimals in the dead zones of disks.

Moreover, when the instability is driven by the relative rotation between the DM and the disk, it acts as a drag on the particles in the disk, potentially facilitating angular momentum transport and thus enhancing the accretion rate and speeding up the formation of planetary cores. We leave a quantitative study of the impacts of our new instability on the evolution of protoplanetary disks and planetary formation to future work.

\subsubsection{Stars}
\label{sec:stars}
Main sequence stars, like the Sun, have densities of about $1.4\,\mathrm{g\,cm^{-3}}$, and thus we anticipate a linear growth time of stellar pulsations due to DM streams of approximately $7$~kyr. More compact stellar remnants have higher densities and correspondingly shorter growth times: neutron stars with typical densities of $\sim 10^{15}\,\mathrm{g\,cm^{-3}}$ have instability growth times of $15$~yr.
We assumed a finite extent (diameter) for Sun‑like stars, white dwarfs, and neutron stars of $L = 1.4 \times 10^6$ km, $14,\!000$ km, and $20$ km, respectively. 
We would therefore expect pulsations to be excited to nonlinear amplitudes in these stars via our new instability. These oscillations range from low frequencies ($\sim2.5$~mHz) in stars like our Sun to very high frequencies ($\sim20$~kHz) in neutron stars.

For neutron stars, the predicted $\sim20$~kHz pulsations lie in a frequency regime that is relevant for pulsar timing array (PTA) experiments. In such experiments, a limiting source of noise on short timescales is ``jitter'', stochastic variations in pulse phase and amplitude arising from single‑pulse variability \citep{Lam2019}. As radio telescopes achieve higher sensitivity, jitter becomes the dominant uncertainty for many millisecond pulsars \citep{Lam2019}.
The instability drives coherent oscillations at $\sim20$\, kHz, which will modulate the arrival times of individual pulses and contribute to the measured jitter. Thus, dedicated searches for narrow‑band signals in the jitter power spectrum could provide a novel observational test of the resonant instability and simultaneously improve our understanding of the noise budget in PTAs.

Moreover, as discussed in Section~\ref{sec:ice}, this leaves potentially observable helioseismic signatures on the Sun, driven on timescales much shorter than its age. We leave quantitative calculations of this effect and the expected amplitudes driven in various types of stars to future work.

\subsubsection{Molecular clouds in galaxies}

Molecular clouds are the coldest ($\sim10$ K), densest ($10^{2}$--$10^{6}$~cm$^{-3}$), and most massive structures in the interstellar medium (ISM). In the Milky Way, they are predominantly found as \textit{Giant Molecular Clouds} (GMCs) with masses of $10^{4}$--$10^{7}$~M$_{\odot}$ and sizes of $L = 20$-$200$\,pc. Composed primarily of molecular hydrogen (H$_2$) and helium, they also host a rich array of complex molecules. Shielded by dust from destructive ultraviolet radiation, their interiors provide the unique conditions necessary for gravitational collapse and star formation.
Using typical cloud densities, the timescale for driven sound waves is about 1--10 Myr. Thus, the relative drift should drive sound waves on scales below the Jeans length of molecular clouds, on a timescale comparable to or smaller than the cloud age ($\sim2$--30~Myr; \citealp{Semadeni+2018}).

Including this instability can therefore impact the star formation rate in simulations of molecular cloud formation. The instability launches sound waves that propagate through the clouds, potentially lowering the star formation rate. These sound waves also provide a source of pressure and drive turbulence within GMCs, which is critical for supporting the clouds against gravitational collapse \citep{Larson:1981}. It is well known that such turbulence exists and must be driven. Typical explanations invoke stirring on small scales (e.g., by stars) or large‑scale driving (e.g., by galactic rotation), but how energy is transmitted into the cloud interior remains poorly understood \citep[see, e.g.,][and references therein]{McKeeOstriker:2007}. The resonant instability presented here offers a novel, self‑consistent mechanism for driving turbulence directly within GMCs, without relying on external or circular arguments. We leave a quantitative study of this effect on MCs to future work.

\subsubsection{Galaxies: Spiral arms}
\label{sec:galaxies}

For typical interstellar conditions ($\rho_{\mathrm{b}} = 1\,m_p/
\mathrm{cm}^3$), a relative drift of $0.4c_s \sim 4$~km~s$^{-1}$ drives density waves on a timescale of $\approx31$~Myr – much shorter than galactic ages ($\sim10^4$~Myr) \citep{Keenan+2014, Naidu+2021}. The fastest mode has a wavelength of $\sim1.6$~kpc, and shorter wavelength modes grow on longer timescales. This applies to a galaxy with a disk diameter of $L = 50$~kpc, as used in Equation~\eqref{eq:urmin}.
However, in a thin galactic disk with differential rotation and a pattern speed $\Omega_{\rm p}$, the standard baryon dispersion relation $\omega^2 = k^2 c_s^2 - 4\pi G \rho_{b,0}$ is replaced by the disk‑specific form $\omega^2 = k^2 c_s^2 + 4\Omega_{\rm p}^2 - 2\pi G \Sigma_0 |k|$, where $\Sigma_0$ is the surface density of the disk \citep{Toomre1964, Binney+Tremaine_08}. Consequently, the resonance mode $k_\mathrm{res}$ (Equation~\eqref{eq:kres}) shifts and depends on the relative rotational speed. Nonetheless, the expected growth time remains short, approximately $30$~Myr. As shown in Table~\ref{tab:param}, the oscillation frequency is very low, giving an evolution timescale of $\sim 2\pi/\omega_{\mathrm{res}} \approx 600$~Myr.

With these caveats, two distinct types of relative motion can exist in galaxies. First, a relative drift between the centers of the DM halo and the galactic disk. In such a case, the instability drives multi‑crescent structures in the galactic disk that are almost perpendicular to the direction of the relative motion. Second, a relative rotational velocity between the galaxy and the DM halo.

Before elaborating on the impacts of relative rotation, we note that DM inside galaxies is kinematically heated.
Our calculations in Appendices \ref{sec:hDMnodrift}\,and\,\ref{app:hdrft} show that the same resonant instability exists in such a case, and the resonance occurs between baryon modes and backward‑propagating DM modes $\omega = k (v_{\mathrm{DM}} \cos \theta - c_{\mathrm{DM}})$.
Whether such a resonance exists in some environment depends on the rotational frequencies of the galactic disk and DM halo, as well as the sound speed of baryons, $c_s$, and velocity dispersion of DM, $c_\mathrm{DM}$. We defer to future work a demonstration that this instability naturally explains constant‑pitch‑angle spiral arms \citep{Grand2013, Savchenko+2013} in disk galaxies and their absence in elliptical galaxies (Broderick \& Shalaby, in preparation).

Both crescent‑like and angular (arm) modes can impact matter accretion and angular momentum transport in the galaxy and may also facilitate accretion onto the central black hole by providing an efficient angular momentum transport mechanism.

The instability may also influence early galaxy formation. Observations from the James Webb Space Telescope (JWST) have revealed surprisingly mature, massive galaxies at high redshifts ($z \gtrsim 10$) with diverse morphologies, including both disk‑like and irregular structures \citep{Labbe2023, Casey2024}. As shown in Section~\ref{sec:universe} and Figure~\ref{fig:cosmology}, the relative drift enhances baryon perturbations through this instability at early epochs, potentially contributing to the rapid assembly and morphological diversity seen in high‑redshift galaxies. The interplay between the resonant instability and the dynamical evolution of early galaxies offers a promising avenue for future investigation.

\subsubsection{Clusters of galaxies}

Galaxy clusters are the universe's most massive gravitationally bound structures, containing hundreds to thousands of galaxies~\citep{miyatake2025cosmology}. They reside at the intersections of the cosmic web, the large-scale filamentary structure that defines the organization of matter in the universe.
A typical cluster spans 1–5 Mpc in diameter and has a total mass of \(10^{14}\)–\(10^{15}\,\mathrm{M}_{\odot}\). The visible galaxies contribute only about 1\% of the cluster's mass. The remainder consists of two components: a hot, X-ray emitting plasma known as the intracluster medium (ICM), which comprises roughly 9\% of the mass, and the dominant dark matter, which constitutes the remaining 90\%. Individual galaxies within clusters move rapidly, with a velocity dispersion of \(800\)–\(1000\,\mathrm{km}\,\mathrm{s}^{-1}\)~\citep{NFW-1996}.

The central regions of galaxy clusters are X-ray bright but are expected to cool radiatively on timescales of a few hundred million years – much shorter than their age (\(\gtrsim10\) Gyr). This discrepancy is known as the cooling flow problem~\citep{Fabian1994, McNamara+2007}. The standard solution invokes feedback from active galactic nuclei (AGN): the supermassive black hole at the cluster center injects energy through jets and outflows, creating buoyant bubbles that dissipate energy as sound waves, heating the surrounding gas~\citep{Fabian2012, McNamaraNulsen2012}. 
It is now well established that AGN jets carry more than sufficient energy to stave off catastrophic cooling in cool-core clusters~\citep{blazariii}. The heat is transported through the ICM via various channels, including turbulent dissipation driven by AGN activity or cluster mergers, turbulent mixing of hot and cold phases, and the dissipation of Alfvén waves excited by streaming cosmic rays (CRs)~\citep{Pfrommer2013, Jacob2017a, Jacob2017b, Ehlert2018}. Thermal conduction from the hot outer regions may also contribute, though simulations indicate it is often insufficient because magnetic fields suppress its efficiency, particularly in cool-core clusters~\citep{ZakamskaNarayan2003, Fabian2012}. Despite the many proposed heating mechanisms, the question of which process dominates in any given cluster remains a topic of active research.

{\it Impact of the new instability.}
There are multiple reasons to expect significant DM–baryon drift velocities in galaxy clusters. At decoupling, the relative drift between DM and baryons is about $5\,c_s$, falling to $\sim2\,c_s$ by $z=0$ (see Appendix~\ref{sec:lincosm}).
Moreover, the motion of constituent galaxies stirs the ICM, generating roughly virialized bulk flows with drift velocities of up to $c_s$. Active galactic nucleus (AGN) feedback can also produce outflows over a wide range of speeds, from large‑scale circulations driven by buoyantly rising bubbles ($0.2c_s$–$c_s$) to the fast interiors of AGN jets themselves (up to $\approx10^3c_s$). Thus, a broad spectrum of relative drifts is expected to be present in the ICM of galaxy clusters.

For a drift velocity of order $c_s \approx 10^{-3}c$ (corresponding to $u_r \sim 3.5$), the instability drives sound waves in the ICM on a timescale of approximately $433$\,Myr with a wavelength of $\sim700$\,kpc (see Table~\ref{tab:param}). This growth time is significantly shorter than the typical cluster age ($\sim10$\,Gyr) and, crucially, is comparable to the radiative cooling time. Since $R=3$, modes with larger $u_r$ grow on similar timescales: for $u_r = 11$, $T_{\mathrm{res}} = 436$\,Myr and $\lambda_{\mathrm{res}} \sim 230$\,kpc.

The ICM is fully ionized, so the driven sound waves are subject to dissipation via ion Landau damping. Ion Landau damping is most efficient on scales comparable to the ion skin depth \citep{LD_2022}, which is much smaller than the driven sound waves. Using Equation (3.2) of \citep{LD_2022}, the damping rate (for the case of electron-ion plasma in thermal equilibrium) is $\Gamma_{\mathrm{LD}} \sim 0.8 k c_s = 0.8 \,\OB \sqrt{1+u_r^2}$.
Hence, the damping time due to ion-landau damping is $T_{\mathrm{LD}} = 1/\Gamma_{\mathrm{LD}} = 1.25 / (\OB\sqrt{1+u_r^2})$.

For the mode with $\lambda_\mathrm{res} \sim 700$ kpc ($u_r \sim 3.5$), the ion-landau damping time is $\sim450$ Myr, comparable to its growth time ($433$ Myr). For the shorter mode with $\lambda_\mathrm{res} \sim 230$ kpc ($u_r = 11$), the damping time is $\sim150$ Myr, while its growth time is $436$ Myr; thus, this mode does not have sufficient time to grow before being damped.

For the mode with $700$ kpc wavelength, the near equality of growth and damping timescales leads to a steady state in which the kinetic energy of the relative drift is efficiently converted into direct heating of the ICM. We therefore conclude that the resonant instability can provide a continuous source of heating that may contribute to a resolution of the cooling flow problem in galaxy clusters.

We note here that the value $700$ kpc is partially a consequence of assuming that the maximum size for a cluster is $L \sim 1$ Mpc; for smaller clusters, the resonant wavemode that sets the equilibrium will be smaller, and vise versa. We defer further quantitative demonstrations to future work.

\subsubsection{Universe}
\label{sec:universe}

In the computation above, we assumed a non‑expanding background, where first‑order Fourier modes for density evolve according to a harmonic oscillator equation. In that case, the dispersion relation determines whether oscillations are stable or exponentially growing.
On cosmological scales, however, the background expands according to the Friedmann–Lemaître–Robertson–Walker (FLRW) metric, and the sound speed also evolves with time. Consequently, the Fourier modes of perturbations obey a damped harmonic oscillator equation (see Appendix~\ref{sec:lincosm} for the full linearized equations). Because of the time dependence of the coefficients, a Laplace transform in time cannot be applied. Therefore, to find solutions, one must solve the evolution equations for the linearized coupled system directly.

We evolve the coupled baryon–dark matter perturbations \eqref{eq:lincosm} from redshift \(z = 1100\), at which the DM drift speed is 5 times larger than the baryon sound speed~\citep{Tseliakhovich+Hirata_2010}. This relative drift speed decreases as the universe expands and scales with the inverse of the cosmological scale factor \(a\), i.e., \(v_{\rm DM} \propto 1/a\) (see also Appendix~\ref{sec:lincosm}). We initialize the perturbations using the Boltzmann solver CLASS \citep{class_2011}. In Appendix~\ref{sec:lincosm}, we provide further details on the assumed thermal history of baryons across different cosmological epochs.

\begin{figure}
\includegraphics[width=1.0\linewidth]{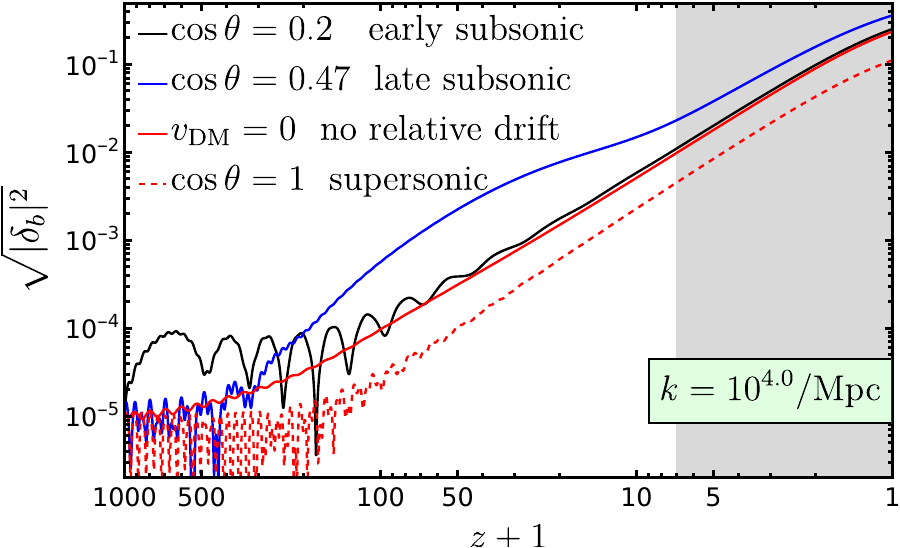}
\caption{\label{fig:cosmology}%
Cosmological evolution of baryon density perturbation $\delta_b$ with $k=10^4\,\mathrm{Mpc}^{-1})$. 
Colors indicate: no drift in solid red, modes along DM streaming direction: dashed red (\(\cos\theta=1\), supersonic drift, density growth suppression), early subsonic drift at recombination (black, \(\cos\theta=1/5\), early density amplification), 
late subsonic drift: \(z\gtrsim200\), blue (\(\cos\theta=0.47\), extra amplification at late times). 
subsonic projected drifts enhance the growth of baryon perturbations while supersonic drifts suppress them.
In the shaded region ($z<6$), the parameterization of the baryon temperature used in Equation~\eqref{eq:Tb} deviates significantly from the much hotter temperatures expected during reionization. However, this increase in $T_b$ corresponds to a decrease in $u_r$ while in a regime where the growth rate is nearly independent of $u_r$ (see right panel of Figure~\ref{fig:urdep}).}
\end{figure}

The Jeans scale changes with cosmic time. It is such that\footnote{We note here that $\Omega_{\mathrm{b},0}$ in this context refers to the ratio of the baryon background density to the critical density at the current epoch and should not be confused with the baryon gravitational frequency $\Omega_{\mathrm{b}}$ defined in previous sections.} $ k_J^2 =  3 \Omega_{\rm b,0}/2 a H_0^2 c_s^2$ 
which means that \(k_J \sim 100\,\mathrm{Mpc}^{-1}\) for \(z \in [1000,200]\), and it increases linearly at lower redshifts, reaching \(k_J = 10^3\,\mathrm{Mpc}^{-1}\) at \(z=0\). Since the resonance occurs only at \(k > k_J\), we show the impact of the resonance on the growth of baryon perturbation for the wavemode  \(k = 10^4\,\mathrm{Mpc}^{-1}\).
We also note that the resonant amplifications shown in Figure~\ref{fig:cosmology} are slightly lower for smaller (but with $k>k_J$) values and higher for larger \(k\).

Figure~\ref{fig:cosmology} shows the cosmological evolution of the baryon density perturbation \(\delta_b \equiv (\rho_{\rm b} - \rho_{{\rm b},0})/\rho_{{\rm b},0}\) for a wave mode with \(k = 10^4\,\mathrm{Mpc}^{-1}\). The evolution when the relative drift is not included (\(v_{\rm DM} = 0\)) is shown by the solid red curve (growth driven by unstable CDM, where \(\delta_b \propto (1+z)^{-1.75}\)).
The evolution for the same mode, when aligned with the dark matter streaming direction (\(\cos\theta = 1\)), which is always supersonic, is shown by the dashed red curve (see Appendix~\ref{sec:lincosm} for when various angles become sonic and subsequently subsonic). In this case, we observe a strong suppression of the baryon perturbation, consistent with the calculations of~\citet{Tseliakhovich+Hirata_2010}.

However, things change significantly at oblique angles. When \(\cos\theta = 1/5\), the dark matter drift is subsonic at (\(z < 1000\)), and our instability can efficiently grow the perturbation. Indeed, at high redshifts (\(z \sim 1000\)), there is a large amplification of the density perturbations. At intermediate obliquities, an example of which is shown by the blue curve (\(\cos\theta = 0.47\)), the projected dark matter speed is subsonic for \(z \lesssim 200\), and at subsequent epochs, i.e., $z < 200$, we see amplification of the baryon density perturbations driven by the resonant instability. The growth driven by the DM–baryon drift is in addition to that normally attributed to cold dark matter modes alone, which are unstable at all scales. Their growth leads to an increase in the gravitational potential at these modes, resulting in the enhanced amplification observed at \(z > 100\) in all cases
\footnote{%
We note that, in the shaded region ($z<6$) of Figure~\ref{fig:cosmology}, the parameterization of the baryon temperature used in Equation~\eqref{eq:Tb} deviates significantly from the much hotter temperatures expected during reionization. However, this increase in $T_b$ corresponds to a decrease in $u_r$ into a regime where the growth rate is nearly independent of $u_r$ (see right panel of Figure~\ref{fig:urdep}). Thus, we expect the lines in this era to still correctly represent the qualitative behavior.
}
.
For angles with subsonic drifts, a similar trend, with even higher enhancement in the density perturbations due to the resonance, is seen at shorter wavelengths (higher \(k\)) compared to the amplifications shown in Figure~\ref{fig:cosmology}.
The resonant enhancement of baryonic perturbations when the relative velocity approaches the sound speed was noted previously by \citet{Yacine+2014}.
They interpreted this as a purely acoustic effect occurring only at $v_{\mathrm{DM}} \gtrsim c_s$ (their Figure 4). In contrast, we have shown that the resonance exists for any subsonic projected drift ($v_{\mathrm{DM}}\cos\theta \le c_s$) and that it drives a genuine gravitational instability, which is more general than the acoustic resonant amplification they described.

Despite assuming both baryons and DM to be collisionless, our instability necessarily dynamically couples the two components to each other.  The energy and momentum in the excited sound waves arise from the relative motions and therefore result in a novel mechanism for momentum exchange between baryons and DM, which may be modeled as an effective non-viscous, collisionless drag.
The instability can produce more than an order‑of‑magnitude enhancement of perturbations in both the velocity gradient and density (see Figure~\ref{fig:cosmology}). Consequently, the backreaction of these perturbations on the background density evolution, particularly at $z<100$, can be substantial.

While we leave a quantitative study for future work, this effect could impact both the Hubble tension \citep{Riess+2016,Hu2023,Verde2024} and the $\sigma_8$ tension \citep{Poulin2023,Amon2022,Pantos2026} through such backreaction and non‑viscous drag, respectively.
Moreover, if these enhanced driven sound waves dissipate energy at such high $z$, this could impact the ionization fraction before the epoch of reionization at $z \approx 7.7$ \citep{Planck+2020}.

The strong density enhancements produced by our instability may play a direct role in the assembly of the first stars~\citep{Klessen+2023}.
The instability naturally generates compact, overdense regions that can serve as cradles for Population III star clusters. Work by \citet{Williams+2026} demonstrates that even when supersonic streaming velocities are present—conditions that usually hinder star formation—Pop III clusters can still reach remarkably high stellar densities, provided that feedback remains weak. By fostering precisely these kinds of dense gas pockets at early cosmic times, the instability we identify could enable the feedback‑free collapse needed to build massive, tightly bound star clusters. These objects would fall within the detection capabilities of facilities such as JWST.

In summary, when we include the effect of expanding spacetime (FLRW cosmology), the relative drift between baryons and dark matter enhances perturbations when they are subsonic and suppresses them when they are supersonic. Moreover, the associated momentum exchange may have significant implications for the background expansion history and the growth of structure, potentially offering a novel resolution to the Hubble and $\sigma_8$ tensions and influencing the thermal history of the universe.

\subsection{Regimes of Inapplicability: Breakdown of Assumptions}

\label{sec:applicability}

This resonantly driven instability can, in principle, drive sound waves on all astrophysical scales since a relative drift between baryons and DM is present at all epochs after recombination \citep{Tseliakhovich+Hirata_2010}. However, the instability relies on several key assumptions; therefore, it is important to discuss various ways in which its impact may not be observed in a realistic environment. The main assumptions are: (i) the existence of cold DM with a well‑defined relative drift, (ii) a sufficiently long coherence time for the DM stream, (iii) a growth time shorter than the age of the system, and (iv) a sufficient number of DM particles within a wavelength to justify a fluid description. When these conditions are not met, the instability may be suppressed or may fail altogether.
The most obvious violation of our assumptions--though cosmologically unlikely--is that DM obeys fundamentally different evolutionary equations than those studied in Appendices~\ref{app:disper} and~\ref{sec:lincosm}, or even the absence of DM altogether.
Below, we discuss further the implicit assumptions that can impact the validity of our new instability.

{\it Coherence requirements.} The instability requires that the DM stream remain coherent across the object of interest for at least one e‑folding time. For ultra‑light DM models (e.g., Bose–Einstein condensates or superfluids), the de Broglie wavelength sets a coherence scale of roughly $1\,\mathrm{kpc}$, corresponding to a coherence time of about $3\,\mathrm{Myr}$ for a typical stream speed of $250\,\mathrm{km\,s^{-1}}$. For weakly interacting massive particles (WIMPs), the quantum coherence scale is much smaller; nevertheless, the primordial streaming velocity field itself remains coherent over $\sim 200\,\mathrm{Mpc}$ \citep{Tseliakhovich+Hirata_2010}. Moreover, inside galaxies, the DM halo breaks into subhalos, producing multiple streams that can excite the same wave mode simultaneously. In such cases, even if an individual stream is coherent for only a short time, the superposition of many streams can maintain an effective growth rate close to the maximum single‑stream value. On cosmological scales, the coherence length at recombination is enormous, and our linear perturbation analysis (Section~\ref{sec:universe}) confirms that the instability enhances baryon density perturbations in the cases we expected.

{\it Granularity of the DM distribution.} A more fundamental requirement is that the DM distribution be effectively continuous on the scale of the resonant wavelength. Our fluid and Vlasov descriptions assume that many DM particles contribute coherently; i.e., the number of DM particles within a volume $A \lambda_{\mathrm{res}}$ (where $A$ is the cross‑sectional area of the system and $\lambda_{\mathrm{res}}$ the resonant wavelength) must be large, $n_{\mathrm{DM}} \, A \lambda_{\mathrm{res}} \gg 1$. If this condition fails, the granularity of DM becomes important, and the collective instability may be suppressed because individual particles cannot establish a smooth gravitational potential perturbation.

This consideration is particularly relevant for ultra‑light DM candidates (e.g., axions or fuzzy DM with masses $\sim 10^{-22}$\,eV). For such particles, the de Broglie wavelength can be comparable to or larger than the resonant scale, especially in low‑density environments. In that regime, the DM distribution is not adequately described by a collisionless fluid or Vlasov equation; instead, it behaves as a coherent wave. The resulting wave interference could modify the gravitational response, potentially altering or suppressing the instability. A full treatment would require a Schrödinger–Poisson (or Gross–Pitaevskii) description, which is beyond the scope of this work. Therefore, the predictions presented here are most directly applicable to particle DM candidates with sufficiently small de Broglie wavelengths (e.g., WIMPs), for which the granularity condition is easily satisfied in all environments considered. For ultra‑light DM, the instability may still operate but requires a dedicated analysis.

{\it Damping and dissipation of the driven waves.}
The observable amplitude of the driven sound waves due to our new instability depends on the balance between growth and various damping mechanisms, as well as on nonlinear saturation. The quality factor \(Q = \omega_{\mathrm{res}} / \Gamma_{\mathrm{res}}\) (see Table~\ref{tab:param}) quantifies the number of oscillations during one e‑fold growth. When \(Q \gg 1\), the waves undergo many cycles before reaching significant amplitude, making them more susceptible to weak damping processes (e.g., viscosity, thermal conduction, or Landau damping). Conversely, when \(Q \lesssim 1\), the waves grow almost monotonically and may reach nonlinear amplitudes before damping becomes important.

Moreover, whether the driven modes are trapped within the environment or propagate outwards also affects their detectability. Using the criterion in Equation~\eqref{eq:prop}, modes with \(\omega_{\mathrm{res}} / \Gamma_{\mathrm{res}} \gg 1\) propagate well beyond the object's size before growing substantially; they may escape the system and dissipate elsewhere, reducing the local signal unless the boundary of the environment reflects and confines them. Modes with \(\omega_{\mathrm{res}} / \Gamma_{\mathrm{res}} \ll 1\) are confined and can build up large amplitudes in situ, provided that nonlinear saturation does not quench the growth prematurely.

In the absence of strong damping, the nonlinear saturation ultimately determines the steady‑state amplitude of the driven waves. Once the wave amplitude becomes large enough, processes such as wave steepening, mode‑mode coupling, or turbulent cascade can balance the linear growth, establishing a saturated amplitude that depends sensitively on the environment. A detailed assessment of saturation requires numerical simulations tailored to each environment and is left for future work. Nevertheless, the existence of the linear instability guarantees a continuous source of wave energy, even if the saturated amplitude is modest.

{\it Marginal regimes.} A borderline case occurs when the growth time is comparable to the coherence time, as in some dense molecular clouds (see Table~\ref{tab:param}); there, the instability may still operate but at a reduced rate. Excluding such marginal cases, the only genuine regimes of inapplicability are those where DM is absent, the relative drift is strictly zero, the growth time exceeds both the coherence time and the age of the system, or the DM particle granularity prevents a collective gravitational response. These conditions are not realized in the standard $\Lambda$CDM paradigm for most astrophysical environments we considered, except possibly for ultra‑light DM candidates, where further study is required.

\section{Potential observable}
\label{sec:obervables}

In this section, we discuss how detecting the DM-driven sound waves can present a novel (in)direct probe for DM.

\subsection{Direct measurements}
\label{sec:direct}

Here, we propose a direct detection of this effect by measuring DM‑driven (vibrational) waves in elastic solids.
In solids, the fluid definition of sound speed using the first‑order perturbation $\nabla^2 p_1 = c_s^2 \nabla^2 \rho_1$ is replaced by $\nabla^2 p_1 = c_1^2 \nabla^2 \rho_1$ (see discussion in Appendix~\ref{app:elasticsolids}). The solid vibrational (dilatational wave) propagation speed $c_1$ is given by \citep{Billingham+Waves}
\[
c_1^2 = \frac{\lambda + 2\mu}{\rho_0},
\]
where $\rho_0$ is the undisturbed density of the elastic solid, and $\lambda$ and $\mu$ are Lamé constants. These are related to Young's modulus $E$ and the dimensionless Poisson's ratio $\nu$ (quantities that can be measured directly for various solids) by
\[
\lambda = \frac{E\nu}{(1+\nu)(1-2\nu)}, \qquad
\mu = \frac{E}{2(1+\nu)}.
\]
All quantities $E$, $\lambda$, and $\mu$ have the same units as pressure (force per unit area). The growth rates given in Equation~\eqref{Eq:grate} apply here with $v_r = v_{\rm DM}\cos\theta / c_1$, where we replace the baryon sound speed with the material sound speed $c_1$ in the definition of $v_r$.

\textit{Possible but a difficult experiment.}
As a concrete—though experimentally challenging—example, consider tungsten. Its mass density is $\rho_b = 19.254\,\mathrm{g\,cm^{-3}}$, Young's modulus is $E = 411\,\mathrm{GPa}$, and Poisson's ratio is $\nu = 0.28$. These yield a vibrational speed of $c_1 = 5.2\,\mathrm{km\,s^{-1}}$. The corresponding Jeans wavelength,
\[
\lambda_J = \frac{2\pi c_1}{\sqrt{4\pi G \rho_b}} = 8.2\times10^3 
\,\mathrm{km},
\]
sets the characteristic scale for gravitational instability.
Depending on the projected dark matter speed, the most unstable wavelength will be a fraction of this length.

For DM with solar neighborhood density $\rho_{\rm DM} = 0.43\,\mathrm{GeV}/c^2\,\mathrm{cm}^3$, the value of $R \sim 2\times10^{-13}$. For modes such that $v_r = 0.1\sqrt{R/9} = 7.4\times10^{-8}$ (i.e., modes almost perpendicular to the DM streaming direction), the growth time is about $17.8$ years; however, the resonance length for such wave modes is extremely long, about $8,200$ km.
On the other hand, for wave modes with $u_r = v_r/\sqrt{1-v_r^2} = 70$, the resonant wavelength is about $115$ km, but the growth time is much longer, approximately $14,\!000$ years. Such values of $u_r$ are easily reachable because the typical DM speed in the solar neighborhood is $v_{\rm DM}\sim220\,\mathrm{km\,s^{-1}} \gg c_1$. Thus, these very slowly growing shorter‑wavelength modes are highly oblique with respect to the DM streaming direction.

In summary, while the tungsten example illustrates the challenges, it may be possible to engineer an apparatus with a much lower sound speed, which would shrink $\lambda_J$ and consequently the resonant wavelength. However, decreasing the growth time remains very difficult. Nonetheless, this intuition motivates us to explore other environments where our new instability can drive sound waves. Their detection could thus serve as a unique method for dark matter detection.

\subsection{Continental Glacier}
\label{sec:ice}
Continental glaciers, also known as ice sheets, are vast, continuous masses of ice that cover large areas of land, exceeding \(50,000\,\text{km}^2\) in extent. Today, they are found only in Greenland and Antarctica, where they comprise approximately 99\% of the world's glacial ice and store about 68\% of Earth's fresh water~\citep{bcotypes_2021}. The Antarctic Ice Sheet, the largest on Earth, has an average thickness of over \(2\,\text{km}\) and reaches up to \(4\,\text{km}\) in places.

Taking a typical ice density of \(\rho \approx 1\,\mathrm{g\,cm^{-3}}\), adopting the solar neighborhood dark matter density, we find that \(R \sim 8.7 \times 10^{-13}\) 
We take a typical size of $L = 2000$\,km and find that the fastest-growing waves are driven nearly perpendicular to the dark matter streaming direction. This arises because the sound speed in ice is approximately \(3.5\,\mathrm{km\,s^{-1}}\), while the expected dark matter speed is \(\sim 230\,\mathrm{km\,s^{-1}}\).
Consequently, the fastest driven sound waves are almost perpendicular to the local streaming direction of dark matter speed.
The dominant mode has a wavelength of approximately \(1400\,\mathrm{km}\), a growth timescale of \(14\)\,kyr, and a frequency of $\sim 16$ mHz.
In contrast, shorter wavelength sound waves with a wavelength of \(100\,\mathrm{km}\) exhibit a longer growth time of $34.2$ kyr and have a higher frequency of $\sim 200$ mHz.

The \(100\,\mathrm{km}\) wavelength is orders of magnitude smaller than the extent of the glaciers, and the driving timescale is negligible compared to the typical glacier age of a few million years. The new instability will drive these detectable sound waves (vibrations) within the Antarctic Ice Sheet.

\subsection{Seismology on Earth, Moon, and Sun}
\label{sec:seismology}

Seismology offers a powerful means to probe the interior structure of planetary bodies and stars. The resonant instability discussed in this work predicts that the relative drift between dark matter and baryons drives coherent sound waves in these objects. The characteristic frequencies and growth timescales, derived in Section~\ref{sec:tscale}, suggest that such waves may be detectable with existing or future seismic networks, providing a novel probe of dark matter.
For typical Earth and Moon parameters, the oscillation frequency of the fastest‑growing mode is about $2\ \mathrm{mHz}$, while for the Sun it is approximately $2.5\ \mathrm{mHz}$, with a wavelength smaller than the extent of these celestial bodies. 

However, several caveats should be kept in mind. First, our calculation assumes a homogeneous, non‑stratified medium; real planets and stars exhibit radial density and pressure gradients that modify the dispersion relation and can couple to different wave modes. Second, the predicted amplitudes of the driven vibrations depend on the efficiency of energy transfer from the DM stream and on damping mechanisms (e.g., viscosity, radiative losses), which we have not modeled, and may result in quality factors well below those listed in Table~\ref{tab:param}. Third, the excited waves may manifest as either body waves (compressional or shear) or surface waves, depending on the depth of the resonance relative to the object’s size; our plane‑wave analysis does not distinguish between these. Finally, the resonance condition can be satisfied for a range of wavenumbers $k$ and angles $\theta$, leading to a spectrum of possible frequencies. For Earth, the Moon, and the Sun, the predicted frequencies lie in the millihertz band ($10^{-3}$–$10^{-2}\ \mathrm{Hz}$); these frequencies overlap with the known helioseismic $p$-mode spectrum. The actual detectability will depend on the amplitudes, which we leave to future work.

\textbf{Earth.} As estimated in Table~\ref{tab:param} and Section~\ref{sec:planets}, the fastest growth time for Earth is about $20$ years for modes that maximize the growth rate. The corresponding resonant wavelength depends on the angle between the wave mode and the DM stream; it can range from scales slightly smaller than the Jeans scale ($\sim 6500$ km, which is approximately equal to the radius of the Earth) down to much smaller values when the projected drift is nearly sonic. Waves at the longer end of this spectrum would be global in extent and could manifest as low-frequency oscillations of the Earth's interior. Modern seismic networks, such as the Global Seismographic Network (GSN), continuously record ground motion with high sensitivity \citep{GSNcitation}. A dedicated search for narrow-band signals at the predicted frequencies could, in principle, reveal the presence of DM-driven vibrations.

\textbf{Moon.} The Moon presents a particularly quiet seismic environment, with negligible atmospheric noise and low tectonic activity. Although lunar seismology data are limited to the Apollo Passive Seismic Experiment \citep{Latham1970, NASA_PSE}, future missions, e.g., Artemis IV \citep{Artemis3_2024}, are expected to deploy new seismometers, including the Lunar Environment Monitoring Station (LEMS) \citep{Marusiak2024}. The lunar density ($\sim3.3\ \mathrm{g\,cm^{-3}}$) and compressional wave speed ($\sim6\ \mathrm{km\,s^{-1}}$) are comparable to Earth's, leading to growth timescales on the order of decades as well. The absence of a thick atmosphere and ocean, combined with the Moon's small size, makes it an attractive target for detecting the predicted seismic signature in the form of low-frequency vibrational oscillations \citep{Schmelzbach2020}.

\textbf{Sun.}~%
Helioseismology has provided exquisite measurements of solar oscillations, revealing a rich spectrum of acoustic modes \citep{Harvey1996, Kosovichev+2025}. From Table~\ref{tab:param}, the characteristic growth time for solar-like densities is $\sim7$~kyr, which is long compared to the solar cycle but still short compared to the age of the Sun. On this timescale, the relative drift drives sound waves with a wavelength of about $10^6$~km and a frequency of $\sim2.5$~mHz. The continuous forcing by the DM–baryon relative drift produces a persistent signal at $2.5$~mHz, while higher frequencies (e.g., $25$~mHz) are driven on longer timescales (e.g., $14.5$~kyr) with correspondingly shorter wavelengths ($\sim1.12\times10^3$~km).

The predicted $\sim2.5$~mHz signal lies well below the solar acoustic cutoff frequency, which is approximately $5.3$~mHz in the quiet Sun \citep{Lamb1909, Jimenez2011}. This cutoff, determined by the density scale height in the solar atmosphere ($\nu_{\rm ac} = c_s/2H_\rho$), sets the highest frequency for trapped acoustic eigenmodes (p‑modes); waves below $\nu_{\rm ac}$ are confined in resonant cavities, while those above propagate into the chromosphere and corona \citep{Jimenez2011, Kosovichev+2025}. Observational estimates of $\nu_{\rm ac}$ range between $5.1$ and $5.7$~mHz \citep{Fossat1992}.

The solar p‑mode power spectrum is dominated by oscillations below $\nu_{\rm ac}$, with power decaying sharply above it. Nevertheless, significant oscillatory power is still observed above $\nu_{\rm ac}$ in the form of pseudomodes (interpreted as interference of traveling waves) and power‑law tails that may indicate a turbulent cascade from oscillations near the cutoff.
The origin of some of this high‑frequency power remains debated \citep{Gizon+2005,Gizon2010}.
The resonant instability we propose offers an additional source: driven sound waves at frequencies both below and above $\nu_{\rm ac}$ would naturally manifest as enhanced p‑mode power, while higher‑frequency waves (with longer growth times) could contribute to the power already observed above the cutoff \citep{Carlsson+2007}. Thus, the instability provides a new physical mechanism that could explain at least part of the solar oscillation spectrum, especially the persistence of power near and above the acoustic cutoff.

In summary, seismic observations of the Earth, Moon, and Sun offer potential avenues to detect the resonant instability predicted in this work. Successful detection would not only confirm the existence of a relative baryon–DM drift but also provide constraints on the local DM density and velocity distribution. We defer a detailed feasibility study, including forward modeling of the expected seismic amplitudes and signal processing strategies, to future work.

\subsection{Cosmological measurements}
\label{sec:cosm}

As shown in Figure~\ref{fig:cosmology}, the instability greatly enhances the baryon 
density perturbation at various cosmological epochs depending on when the projected relative drift becomes subsonic. 
This high redshift amplification occurs in the linear regime, i.e., where $\delta_b \ll 1$. Consequently, any amplification due to 
this effect at such scales would not be severely contaminated by nonlinear processes.

Given the mean baryon density, the total background mass enclosed in the wave 
modes we considered, \(k = 10^4\) Mpc\(^{-1}\), i.e, $R = 2 \pi \times 10^{-4}$ Mpc, is decreasing as the universe expands, i.e., $M \propto a^{-3}$.
It ranges from $M = 8 \times 10^8 M_{\odot}$ at $z=500$ to  \(8.5\times 10^3 \, M_{\odot}\) at $z=10$.
This implies that running hydrodynamical cosmological simulations to quantify the 
impact on the distribution and evolution of collapsed structures in the early universe 
would require resolving such mass scales, making it potentially computationally challenging, especially at late times. 

However, we will elaborate on detection strategies for observing such density amplification 
using the 21 cm signal in a forthcoming publication (Shalaby et al., in preparation).

\section{Summary and Conclusions}
\label{sec:sandc}

In this paper, we have uncovered a new fundamental mechanism in the evolution of cosmic structure: a resonant gravitational instability driven by the relative streaming of baryons and dark matter (DM).  This instability is a direct consequence of the standard cosmological mode and does not require any new physics.
We have presented, for the first time, a comprehensive analysis of this instability and find that it fundamentally alters the standard picture of gravitational instability and provides a new lens through which to understand the interaction of dark and visible matter across all scales.

By deriving and analyzing the linear dispersion relation for the coupled baryon-DM system, we identified the key conditions under which this instability operates. The primary results of our analysis can be summarized as follows:
\begin{enumerate}
    \item \textbf{Resonant Instability:} We demonstrated that when the projected relative drift of DM with respect to the baryon sound speed is subsonic ($v_{r} \equiv v_{\mathrm{DM}}\cos\theta / c_s < 1$), a resonance occurs at scales smaller than the baryon Jeans scale. This resonance couples the stable oscillatory modes of the baryons with the Doppler-shifted modes of the DM, leading to a novel, driven instability. The growth rate at this resonance can significantly exceed the intrinsic growth rate of cold dark matter alone.
    \item \textbf{Stabilized Window:} A key feature of this instability, particularly in baryon-dominated environments ($\Omega_{\mathrm{DM}} \ll \Omega_{\mathrm{b}}$), is the emergence of a stable window for both baryon and DM perturbations. This window exists for wave modes between the classical Jeans scale and the new resonant scale, where perturbations remain completely stable in the absence of other driving forces.
    \item \textbf{Supersonic Suppression:} In the opposite regime, where the projected relative drift is supersonic ($v_{r} > 1$), the growth of matter perturbations is substantially suppressed. This effect aligns with previous findings in cosmological contexts.
    
    \item \textbf{Analytical and Numerical Approximations:} We provided an accurate analytical approximation for the growth rate at resonance in Equation~\eqref{Eq:grate}. This expression reveals the dependence on the DM-baryon density ratio $R^2$ and the effective relative drift $u_r = v_r / \sqrt{1-v_r^2}$, providing a powerful tool for estimating the instability's impact across different physical environments.

    \item \textbf{Momentum Exchange and Collisionless Drag:} The instability not only drives sound waves but also transfers momentum between DM and baryons, creating a collisionless non-viscous drag. This effect is generic and can influence the dynamics of baryons and dark matter across all scales, from planets to the universe.

\end{enumerate}

The implications of our findings are potentially far-reaching. By showing that a ubiquitous relative drift can excite sound waves in baryons on timescales comparable to or shorter than the dynamical times of systems ranging from planets to galaxy clusters, we have identified a potential new source of energy injection and structure formation that has been previously overlooked. Specifically:
\begin{itemize}
    \item \textbf{Astrophysical Systems:} The instability can provide a compelling explanation for the persistence of spiral arms in galaxies, offer a potential heating mechanism to resolve the cooling flow problem in galaxy clusters, and predict the existence of characteristic, externally-driven vibrational modes in planets, stars, and continental ice sheets. The latter suggests a novel, albeit challenging, avenue for the direct detection of dark matter through seismology. Moreover, the associated momentum exchange produces a collisionless drag that can modify the bulk dynamics of baryons in any system where the instability is active.

    \item \textbf{Cosmological Evolution:} Our linear cosmological analysis demonstrates that this effect leaves a distinct signature in the linear growth of structure. The dependence of enhancement or suppression on the angle between the wave mode and the DM streaming direction implies that the large‑scale structure of the universe will not be isotropic with respect to the primordial DM flow. This anisotropy may be imprinted on the $21\mathrm{cm}$ signal and the distribution of early galaxies, offering a potential observational test of the CDM paradigm. 
    The associated momentum exchange and collisionless drag can also \textit{backreact on the background expansion, potentially influencing both the growth of structure and the cosmic expansion history.}
    
    \item \textbf{A New Probe of Dark Matter:} The resonant instability turns the relative motion of dark matter from a nuisance parameter into a powerful diagnostic tool. The specific scales and rates at which sound waves are driven are sensitive to the properties of DM, such as its velocity dispersion (i.e., whether it is cold, warm, or hot) and its local density. This opens up a new class of observational probes, from seismic measurements in terrestrial bodies to the analysis of galactic spiral structure and the large-scale clustering of matter in the early universe.
  
\end{itemize}

The growth times, resonant wavelengths, and frequencies quoted for astrophysical systems in \autoref{tab:param} are order-of-magnitude estimates obtained by applying the homogeneous dispersion relation with local parameters.
Establishing whether these modes exist with comparable growth rates in realistic, finite, stratified systems—and whether they reach observable amplitudes—will require system-specific calculations that incorporate the appropriate equilibrium structure, mode spectrum, boundary conditions, damping processes, and the coherence and phase-space distribution of the dark matter.

Our work establishes the theoretical foundation for this effect. Future work will focus on quantitative modeling in specific environments, including high-resolution simulations of galactic disks and protoplanetary disks, detailed predictions for the 21~cm power spectrum, and the development of experimental concepts to detect these DM-driven waves in the solar system. The "sound of the universe," as we have termed it, is not just a metaphor but a potentially observable phenomenon that promises to deepen our understanding of the dark sector and its interplay with the visible cosmos.

\section*{Acknowledgments}
We thank Neal Dalal and Brian McNamara for numerous fruitful discussions following the identification of this instability.
This work was supported in part by Perimeter Institute for Theoretical Physics.
Research at Perimeter Institute is supported by the Government of Canada through the Department of Innovation, Science and Economic Development Canada and by the Province of Ontario through the Ministry of Economic Development, Job Creation and Trade.
M.S. receives additional support through the Horizon AstroPhysics Initiative (HAPI), a joint venture of the University of Waterloo and Perimeter Institute for Theoretical Physics.
A.E.B. receives additional financial support from the Natural Sciences and Engineering Research Council of Canada through a Discovery Grant.

\let\newpage\relax
\bibliography{refs}
\bibliographystyle{aasjournal}
\let\newpage\newpage

\appendix

\section{Derivation for the dispersion relation}
\label{app:disper}

\subsection{Governing equations}

We model the baryons as an ideal fluid, whose evolution is governed by the Euler equations

\begin{eqnarray}
&&
\partial_t \rho_b + \nabla \cdot \rho_b 
\vec{v}_b=0
\label{Eq:density}
\\
&&
\partial_t \rho_b \vec{v}_b + 
\nabla p_b +
\nabla \cdot \left[ \rho_b \vec{v} _b\vec{v}_b ~  \right] 
=
\rho_b \vec{a} (\vec{x}).
\label{Eq:momentum}
\end{eqnarray}
Here, $p$ is the isotropic fluid pressure
, $\rho_b$ is the local mass density of baryons, and $\vec{a} = \vec{a}_{\rm gravity}(\vec{x})$ is the gravitational acceleration. 
Equations~\eqref{Eq:density} and \eqref{Eq:momentum} can be combined into
\begin{eqnarray}
\partial_t \vec{v}_b 
+ \vec{v}_b \cdot \nabla   \vec{v}_b
+ \frac{ \nabla p_b }{ \rho_b  } 
=
\vec{a}.
\label{Eq:velocity}
\end{eqnarray}

For the DM particles, we model their evolution using the Vlasov equation; i.e., the phase space distribution for DM particles, $ f_\mathrm{DM}$, evolves solely due to the gravitational forces it experiences.
\begin{eqnarray}
&&
 \partial_t f_\mathrm{DM} + 
\vec{v} \cdot \nabla   f_\mathrm{DM} 
+
\vec{a} \cdot  \nabla_{\vec{u}}   f_\mathrm{DM}
=
0.
\label{eq:vlasov}
\end{eqnarray}

The gravitational acceleration obeys the Poisson equation; thus, we can write
\begin{eqnarray}
\vec{a} 
=
- \nabla \phi 
\quad
\text{and}
\quad
\nabla^2 \phi =
 4 \pi G ( \rho_b + \rho_\mathrm{DM} ),
\label{Eq:acc}
\end{eqnarray}
where $\rho_b$ ($\rho_\mathrm{DM}$) is the mass density for baryons (DM), and $\phi$ is the gravitational potential.
We note here that the derivative in the last term of equation ~\eqref{eq:vlasov} is with respect to $u = \gamma v $, where $\gamma$ is the Lorentz factor. Below, we consider only non-relativistic cases for which $\vec{u} = \gamma \vec{v} \sim \vec{v}$.

\subsection{Linear perturbation for the coupled system}

To linearize, we assume some equilibrium ($0$) and a small perturbation ($1$) away from the equilibrium configuration.
The equilibrium is such that there is a uniform, stationary, and non-evolving configuration, i.e., $\partial_t n_{\mathrm{b},0} = 0~ \& ~ \partial_t v_{\mathbf{b},0} = 0$. Then, we add the first order perturbation, i.e., we take 
\begin{eqnarray}
&&
\rho_b = \rho_{b,0} + \rho_{b,1}
~~ \& ~~
\vec{v}_b = \vec{v}_{b,0} + \vec{v}_{b,1}
\nonumber \\
&&
p_b = p_{b,0} + p_{b,1}
~~ \& ~~
f_\mathrm{DM} = f_{\mathrm{DM},0} + f_{\mathrm{DM},1}
 \nonumber \\
&&
\Rightarrow
~~
\vec{a} = \vec{a}_0 + \vec{a}_1
~~\text{or}
~~
\phi = \phi_0 + \phi_1
\end{eqnarray}

For the baryons, we linearize equations \eqref{Eq:density} and \eqref{Eq:velocity}, and for the DM, we linearize equation \eqref{eq:vlasov}.
For the force terms, we linearize equation \eqref{Eq:acc}.
We assume that only the linear terms depend on $\vec{x}$ and $t$.

We also make use of the famous Jean's swindle, i.e., We assume $\vec{a}_0 = - \nabla \phi_0 =0$, despite the fact that $\nabla \cdot \vec{a}_0 = - \nabla^2 \phi_0 = 4 \pi G \sum_s \rho_{b,0} \neq 0$.

The linearized equations (in the baryon rest frame, i.e., $\vec{v}_{b,0}=0$) are given by
\begin{eqnarray}
&&
\partial_t \rho_{b,1} +  \rho_{b,0} \nabla \cdot 
\vec{v}_{b,1}=0
\\ && 
\partial_t \vec{v}_{b,1} + 
\frac{ \nabla p_{b,1} }{ \rho_{b,0}  }=
\vec{a}_1 = - \nabla \phi_1
\\ &&
 \partial_t f_{\mathrm{DM},1} + 
\vec{v} \cdot \nabla   f_{\mathrm{DM},1} 
- \nabla \phi_1  \cdot  \nabla_{\vec{u}}   f_{\mathrm{DM},0}
=
0
\\ &&
\nabla^2 \phi_1 = 4 \pi G \left(\rho_{b,1} + m_\mathrm{DM} \int d^3 u f_{\mathrm{DM},1} \right)
\end{eqnarray}
Where $m_\mathrm{DM}$ is the elementary mass of dark matter.

\subsection{Linear Dispersion relation}

In the Fourier domain, the linearized equations are 
    
\begin{eqnarray}
&&
-i \omega \rho_{b,1} + i  \rho_{b,0} \vec{k} \cdot 
\vec{v}_{b,1}=0
  \Rightarrow  
\omega \frac{n_{b,1}}{n_{b,0}} = \vec{k} \cdot 
\vec{v}_{b,1}
\\ && 
-i \omega \vec{v}_{b,1} + 
\frac{ i c_s^2 \vec{k} \rho_{b,1} }{ \rho_{b,0}  }
= - i\vec{k} \phi_1
\nonumber \\
&& \quad \Rightarrow \quad 
-i \omega^2 \frac{\rho_{b,1}}{\rho_{b,0}} + 
i c_s^2 k^2  \frac{ \rho_{b,1} }{ \rho_{b,0}  }
= - i k^2 \phi_1
\\ &&
-i \omega f_{\mathrm{DM},1} + 
i \vec{v} \cdot \vec{k}   f_{\mathrm{DM},1} 
- i \vec{k} \phi_1  \cdot  \nabla_{\vec{v}}   f_{\mathrm{DM},0}
=
0
\nonumber 
\\ && 
\quad \Rightarrow \quad
f_{\mathrm{DM},1} = - \frac{\vec{k}  \cdot  \nabla_{\vec{v}} f_{\mathrm{DM},0} }{\omega - \vec{k} \cdot \vec{v}} \phi_1 
\\ &&
- k^2 \phi_1 = 4 \pi G \left(\rho_{b,1} + m_\mathrm{DM} \int d^3 v f_{\mathrm{DM},1} \right)
\qquad \quad
\end{eqnarray}
Where we define the sound speed as
$ \nabla p_{b,1}  =  c_s^2 \nabla \rho_{b,1}  $, and $c_s$ depends on $p_{b,0}$ and $\rho_{b,0}$.
That is,
\begin{eqnarray}
&&
( \omega^2  - c_s^2 k^2  ) \rho_{b,1}  
=   k^2 \rho_{b,0}  \phi_1 \qquad \& 
 \nonumber \\
\rho_{\mathrm{DM},1} 
&=&
m_\mathrm{DM} \int d^3 v f_{\mathrm{DM},1} 
= - m_\mathrm{DM}  \int d^3 v \frac{\vec{k}  \cdot  \nabla_{\vec{v}} f_{\mathrm{DM},0} }{\omega - \vec{k} \cdot \vec{v}} \phi_1 
 \nonumber \\
&=& m_\mathrm{DM}  k^2 \phi_1  \int \frac{  d^3 v f_{\mathrm{DM},0} }{(\omega - \vec{k} \cdot \vec{v})^2}
\label{eq:disp02}
\end{eqnarray}
The above equation has an important implication; the solution of the dispersion relation studied below gives the oscillatory or growing behavior of $\phi_1$, which is the same behavior as the growth of the initial mass density perturbation in both baryons and DM normalized to their initial mass density; that is, $\rho_{1,b}/\rho_{0,b} \propto \phi_1 $ and $\rho_{1,\mathrm{DM}}/\rho_{0,\mathrm{DM}} \propto \phi_1 $.

For hot/warm DM cases at thermal equilibrium in the non-relativistic regime, the equilibrium velocity distribution is given by
\begin{eqnarray}    
f_{\mathrm{DM},0} =
\frac{n_{\mathrm{DM},0} }{( 2 \pi T_\mathrm{DM} )^{3/2}}
\exp
\left[ - \frac{ m_\mathrm{DM} (\vec{v} - \vec{v}_\mathrm{DM})^2}{2 k_B T_\mathrm{DM} } \right], \qquad
\end{eqnarray}
where $\vec{v}_\mathrm{DM}$ is the DM average relative drift, and $T_\mathrm{DM}$ is the DM temperature.
Therefore, we can write that

\begin{align}
\frac{ \rho_{\mathrm{DM},1}/k^2}{ \rho_\mathrm{DM,0}  \phi_1}
=& 
\int 
\frac{d^3 v }{( 2 \pi T_\mathrm{DM} )^{3/2}} 
\frac{f_{\mathrm{DM},1}}{n_{\mathrm{DM},0}}
 \nonumber \\
 =&
\int 
\frac{d^3 v }{( 2 \pi T_\mathrm{DM} )^{3/2}} 
\frac{  ~ e^{-\frac{ m_\mathrm{DM} (\vec{v} - \vec{v}_\mathrm{DM})^2}{2 k_B T_\mathrm{DM} }} 
}{(\omega - \vec{k} \cdot \vec{v})^2} 
 \nonumber \\
 =&
\int 
\frac{ d^3 \chi }{( 2 \pi T_\mathrm{DM} )^{3/2}} 
\frac{  \quad e^{-\frac{ m_\mathrm{DM} \chi^2}{2 k_B T_\mathrm{DM} }} 
}{
(\omega - \vec{k} \cdot \vec{v}_\mathrm{DM} -  \vec{k} \cdot \mathbf{\chi})^2} 
 \nonumber \\
 =&
\int 
\frac{ d^3 \chi }{( 2 \pi T_\mathrm{DM} )^{3/2}} 
\frac{ \quad e^{-\frac{ m_\mathrm{DM} \chi^2}{2 k_B T_\mathrm{DM} }} 
}{
(\Tilde{\omega} -  \vec{k} \cdot \mathbf{\chi})^2} 
 \nonumber \\
 =&
\frac{1/\Tilde{\omega}^2 }{( 2 \pi T_\mathrm{DM} )^{3/2}} 
\int 
 d^3 \chi 
\frac{ \quad e^{-\frac{ m_\mathrm{DM} \chi^2}{2 k_B T_\mathrm{DM} }} 
}{ (1-  \vec{k} \cdot \mathbf{\chi}/\Tilde{\omega})^2} 
\nonumber \\
 \approx &
\int 
\frac{ d^3 \chi ~ e^{\frac{ m_\mathrm{DM} \chi^2}{2 k_B T_\mathrm{DM} }} }{\Tilde{\omega}^2 ( 2 \pi T_\mathrm{DM} )^{3/2} }
\left[ 1 
+2 \frac{\vec{k} \cdot \mathbf{\chi}}{\Tilde{\omega}  }
+3 \left( \frac{ \vec{k} \cdot \mathbf{\chi}}{\Tilde{\omega} }\right)^2 
\right]
 \nonumber \\
 = &
\frac{ 1 }{\Tilde{\omega}^2}
\left(1 + 3 \frac{ k^2  k_B T_\mathrm{DM} }{ \Tilde{\omega}^2} \right)
\nonumber \\ \approx &
\frac{1 }
{
\Tilde{\omega}^2
\left(1 - 3 k^2  k_B T_\mathrm{DM} /\Tilde{\omega}^2 \right)
}
\nonumber \\
=& 
\frac{1 }
{
\left(\omega - \vec{k} \cdot \vec{v}_\mathrm{DM} \right) ^2 -   k^2  c^2_\mathrm{DM}  
}.
\label{eq:fdm1}
\end{align}

Where, $c_\mathrm{DM}^2 \equiv 3 k_BT_\mathrm{DM}$, is the dark matter velocity dispersion.
In equation~\eqref{eq:fdm1}, we retain only the principal part of the integral. Including the full contributions would introduce the Landau damping phenomenon, which changes the growth rates at wavelengths longer than the Jeans length \citep[e.g., Chapter\,5 of][]{Binney+Tremaine_08}. 
We focus on wavelengths shorter than the Jeans scale, where this simplification yields accurate results.
Then, we can write
\begin{eqnarray}
- k^2 \phi_1 &=& 4 \pi G \left( \rho_{b,1} + m_\mathrm{DM} \int d^3 v f_{\mathrm{DM},1} \right)
\nonumber \\
&=&
4 \pi G \left(\frac{k^2 \rho_{b,0} \phi_1 }{ \omega^2 - k^2 c_s^2}+ 
\frac{k^2 \rho_{\mathrm{DM},0} \phi_1}{
\left(\omega - \vec{k} \cdot \vec{v}_\mathrm{DM} \right) ^2 -   k^2  c^2_\mathrm{DM}  
}\right) 
\nonumber 
\end{eqnarray}
Therefore, the linear dispersion relation is given by
\begin{eqnarray}
-1 &=& 
\frac{4 \pi G  \rho_{b,0} }{ \omega^2 - k^2 c_s^2}+ 
\frac{4 \pi G  \rho_{\mathrm{DM},0}}{
(\omega - \vec{k} \cdot \vec{v}_\mathrm{DM})^2 -   k^2  c^2_\mathrm{DM}  }
\nonumber \\
&=&
\frac{\Omega_b^2 }{ \omega^2 - k^2 c_s^2}+ 
\frac{\Omega_\mathrm{DM}^2}{
(\omega - \vec{k} \cdot \vec{v}_\mathrm{DM})^2 -   k^2  c^2_\mathrm{DM} }
\label{Eq:disp18}
\end{eqnarray}
where we define 
$\Omega_{b}^2 \equiv 4 \pi G  \rho_{b,0}  $ and 
$\Omega_{\mathrm{DM}}^2 \equiv 4 \pi G  \rho_{\mathrm{DM},0}  $.

One can also find the cold limit for DM directly by using Equation~\eqref{eq:disp02} and $f_{\mathrm{DM},0} = n_{\mathrm{DM},0} \delta^3( \vec{v} - \vec{v}_\mathrm{DM})$. Thus, we can write the DM first order perturbation as 
\begin{eqnarray}
\rho_{\mathrm{DM},1} 
=  \frac{ m_\mathrm{DM}  n_{\mathrm{DM},0} k^2 \phi_1}{(\omega - \vec{k} \cdot \vec{v}_\mathrm{DM})^2}
=  \frac{ \rho_{\mathrm{DM},0} k^2 \phi_1}{(\omega - \vec{k} \cdot \vec{v}_\mathrm{DM})^2}
\end{eqnarray}
This is the same contribution to the dispersion relation found when setting $c_\mathrm{DM}=0$ in equation \eqref{Eq:disp18}.

To solve the dispersion relation, we define the following dimensionless variables: $x = \omega/\Omega_\mathrm{b}$, $y = k c_s / \Omega_\mathrm{b}$, $R = \Omega_\mathrm{DM}/\Omega_\mathrm{b}$, $c_r = c_\mathrm{DM}/c_s$, which is the ratio of the DM velocity dispersion to the baryon sound speed, and $v_r = v_\mathrm{DM} \cos(\theta)/c_s$ is the projected DM speed with respect to the sound speed propagation direction, normalized by the baryon sound speed $c_s$. $\theta$ is the angle between  the DM propagation direction and the sound wave propagation direction. Thus, the dispersion relation can be re-written as
\begin{eqnarray}
-1 = 
\frac{1}{ x^2 - y^2}+ 
\frac{R^2}{
(x - y v_r)^2 - y^2 c_r^2}
\label{Eq:disp01}
\end{eqnarray}

Below, we present various solutions for this dispersion relation when there is no relative drift, i.e., $ \vec{v}_\mathrm{DM}=0$ (Appendix~\ref{app:nodrift}), followed by a presentation of the impacts of the relative drift (Appendix~\ref{app:wdrift}).

\section{Impact of  non-drifting DM on Baryons Jeans instability}
\label{app:nodrift}

Here, we study the cases where there is no relative drift between the DM and baryons, i.e, $v_r=0$.
We start with the case of cold DM and then discuss the case of DM with a non-vanishing velocity dispersion.

\subsection{Cold DM Case}

\begin{figure}
\includegraphics[width=1\linewidth]{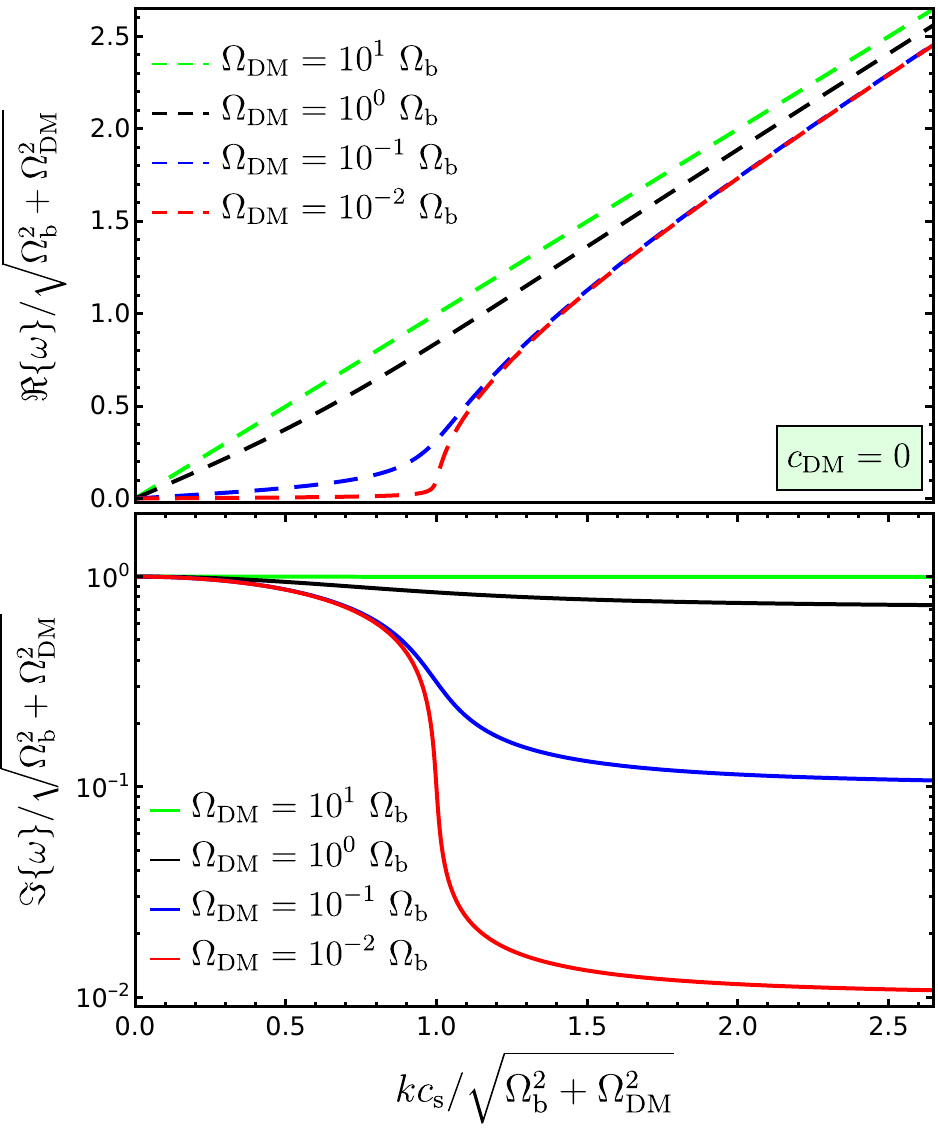}
\caption{\label{fig:novdr}%
The top panel shows the oscillatory behavior, while the bottom panel shows  the growth, of gravitational perturbations in the presence of cold DM ($c_\mathrm{DM}=0$) with various densities.
we emphasize here that the solutions in top and bottom panels are different branches, that is for real solution shown in the top penal, $\Im\{\omega\} =0$, and for the growing solutions shown in the bottom panel $\Re\{\omega\} =0$.
As we show in Appendix~\ref{app:nodrift}, the oscillatory modes (whose oscillation frequencies are shown in the top panel) are completely stable, i.e., these have $\Im\{\omega\} = 0$. Conversely, the unstable modes (whose growth rates are shown in the bottom panel) are non-oscillatory, i.e., for these, $\Re\{\omega\} = 0$. These solution (and their negative values) form all solutions when there is no relative baryon-DM drift.
}
\end{figure}

Here, we investigate the impact of cold DM (i.e., $c_r=0$) 
on the Jeans instability of thermal baryons. In this case, the dispersion relation is
\begin{equation}
-1 = \frac{1}{x^2 - y^2} + \frac{R^2}{x^2}.
\label{Eq:disp00_novdr}
\end{equation}

\begin{figure*}
\includegraphics[width=1\linewidth]{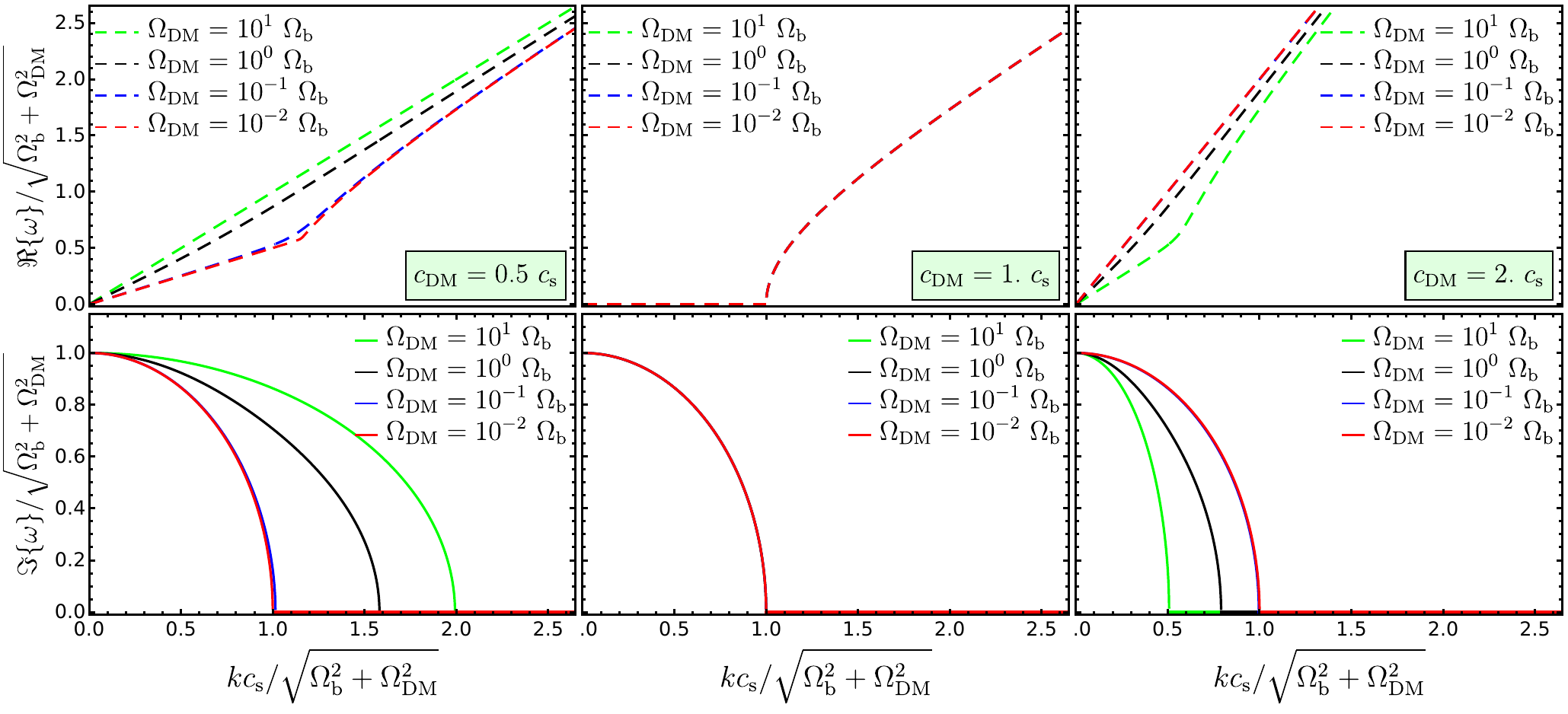}
\caption{\label{fig:novdrCr}%
Same as in Figure~\ref{fig:novdr} but with a non vanishing DM sound speed. These are the solution found from Equation~\eqref{Eq:sol_cDM}.
This show an important effect, when the DM are hot, scales below the effective Jeans scale are completely stable, i.e., with zero growth rates.
This scale in the Baryon Jeans scale in the Baryon dominated case, i.e., $R\ll 1$, and is roughly $k c_\mathrm{DM} = \ODM$ in DM dominated case, i.e., $R \gg 1$.
}
\end{figure*}

This can be rearranged as
\begin{equation}
   2 x^2 = y^2 - 1 - R^2 \pm \sqrt{4 R^2 y^2 + \left(y^2 - 1 - R^2\right)^2}.
   \label{eq:novdr}
\end{equation}
Since $4 R^2 y^2  > 0$, all solutions satisfy $x^2 \in \mathbb{R}$.
That is, all solutions for $x$ (or $\omega$) are purely real or purely imaginary. That is, perturbations are stable and oscillatory when $x^2 > 0$, or unstable (growing or decaying) without oscillation when $x^2 < 0$.
These stable and unstable solutions are shown in the upper and lower panels of Figure~\ref{fig:novdr}, respectively, for various values of $R = \ODM /\OB$.
The solutions in Equation \eqref{eq:novdr} have many interesting limits that are discussed below.

In the long-wavelength limit, i.e., $k \to 0$ or $y = 0$, we find the following.
\begin{eqnarray}
x^2 \approx \left\{ 0 , ~ -(1+R^2) \right\}
\Rightarrow
\omega \approx \left\{0 ,  \pm i \sqrt{\Omega_\mathrm{b}^2 + \Omega_\mathrm{DM}^2} \right\} ~~ \qquad ~~
\end{eqnarray}

In the short-wavelength limit, i.e., $k \gg k_J$ or $y \gg 1$,
\begin{eqnarray}
x^2 \approx \left\{y^2, -R^2 \right\}
~~ \Rightarrow ~~
\omega \approx \left\{ k c_s , \pm i \Omega_\mathrm{DM} \right\}. \qquad 
\end{eqnarray}

Finally, on the Jeans scale of baryons $k = k_J = \OB/c_s$ (i.e., $y=1$), we find
\begin{equation}
2 x^2 = -R^2 \pm R \sqrt{R^2 + 4}.
\end{equation}
When $R=0$, we recover the expected solution $\omega = 0$.
On the other hand, when a non-vanishing DM density is included ($R > 0$), the solutions at the Jeans scale bifurcate. The first solution is stable and purely oscillatory, i.e, $ 2 x^2 = -R^2 + R \sqrt{R^2+4}  > 0$,
with an oscillation frequency 
\begin{equation}
\omega^2 = \frac{ -\Omega_\mathrm{DM}^2 + \Omega_\mathrm{DM} \sqrt{\Omega_\mathrm{DM}^2 + 4 \Omega_\mathrm{b}^2} }{2}.
\end{equation}
It is important to note here that $\Omega_\mathrm{DM}^2 + \Omega_\mathrm{b}^2 = 4 \pi G (\rho_{\mathrm{b},0} + \rho_{\mathrm{DM},0}) $ is the square of the gravitational frequency of the total matter (baryons and DM).
For $R \gg 1$, i.e., $\Omega_\mathrm{DM} \gg \Omega_\mathrm{b}$ this simplifies to $\omega \approx \OB$. This is the upper limit for DM dominated oscillation frequency at the Jeans scale.

The second solution is a purely growing/damping mode,
\begin{eqnarray}
    x^2 &=& -\frac{1}{2} \left( R^2 + R \sqrt{R^2 + 4} \right)
\nonumber \\
 \Rightarrow ~ && \hspace{-0.2cm}
   \Im(\omega) = \pm \sqrt{ \frac{ \Omega_\mathrm{DM}^2 + \Omega_\mathrm{DM} \sqrt{\Omega_\mathrm{DM}^2 + 4 \Omega_\mathrm{b}^2} }{2} }.
\end{eqnarray}
Similarly, for $R \gg 1$, i.e., $\Omega_\mathrm{DM} \gg \Omega_\mathrm{b} ~\Rightarrow~\Im(\omega) \approx \Omega_\mathrm{DM}$.

In summary, Figure~\ref{fig:novdr} demonstrates that for the case of $R \geq 1$ (i.e., $\rho_\mathrm{DM} \geq \rho_\mathrm{b}$), the Jeans instability for gravitational perturbations is fundamentally different from that in the case of $R=0$. Two key changes are evident: firstly, stable wave modes at all scales oscillate with frequency $k c_s$, not only modes much shorter than the Jeans scale, as in the $R=0$ case. Secondly, unstable wave modes grow at a constant rate of $\sqrt{\Omega_\mathrm{DM}^2 + \Omega_\mathrm{b}^2}$ on all scales.

For the case of $R \ll 1$, the Jeans instability is very similar to that of $R=0$ (the baryon-only case), except that on scales shorter than the Jeans scale, the gravitational perturbations are unstable, with a growth rate $ \approx \Omega_\mathrm{DM}$.

These features are important to understand how the resonant instability discussed in Appendix~\ref{app:wdrift} is much more evident in the case of $R \leq 1$.

\subsection{Hot DM Case}
\label{sec:hDMnodrift}

In the case where the DM sound speed is not negligible, the dispersion relation is given by
\begin{equation}
-1 = \frac{1}{x^2 - y^2} + \frac{R^2}{x^2 - y^2 c_r^2}.
\label{Eq:disp00_cDM}
\end{equation}
Again, this can be rearranged as
\begin{eqnarray}
2 x^2 &=& 
y^2 (1+c_r^2) -1 -R^2  \pm
\nonumber \\ &&
\sqrt{
4 R^2 y^2(1-c_r^2) + \left[ y^2 (1-c_r^2) - 1 - R^2 \right]^2 
} 
\qquad \quad
\label{Eq:sol_cDM}
\end{eqnarray}
When $c_r =0$, the above expression reduces to the same expression as in Equation~\eqref{eq:novdr}, and similarly, one can show that the coefficient inside the square root is always positive for any value of $c_r \geq 0$.
Therefore, the solutions $x$ are either purely oscillatory and stable, or are growing without any oscillation.

From the dispersion relation (Equation~\eqref{Eq:disp00_cDM}), one can see that when $c_r =1$, baryons and DM are indistinguishable; thus, all scales longer than the effective Jeans scale are unstable, while shorter scales are stable and oscillatory.
The effective Jeans wave mode is $k c_s/\sqrt{\ODM^2+\OB^2} =1$.
In Figure~\ref{fig:novdrCr} we show the solutions in Equation~\eqref{Eq:sol_cDM} for various values of $R$ and $c_r \equiv c_\mathrm{DM}/c_s$. 
This shows how $c_r \neq 0$ changes the stability of gravitational perturbations compared to the case of $c_r=0$ (Figure~\ref{fig:novdr}).

\section{Driven Resonant instability}
\label{app:wdrift}

Here, we present the solution for the dispersion relation when  the impact of the relative drift between baryons and DM is included for both cases of cold and hot DM.

\subsection{Cold DM case with relative drift}
\label{app:cdrft}

In such a case, the dispersion relation is given by
\begin{eqnarray}
-1 = 
\frac{1}{ x^2 - y^2}+ 
\frac{R^2}{
(x - y v_r)^2}
\label{Eq:disp01}
\end{eqnarray}

\subsubsection{Streaming Resonance Location}

The resonance scale is derived in Section~\ref{sec:resonance}. It is such that $k c_s \sqrt{1 - v_r^2} = \Omega_{\mathrm{b}}$. In dimensionless units, the streaming resonance between the stable branch of baryons, i.e., for $y > 1$, is such that $x^2 = y^2 - 1$, and the real frequency of the unstable, non-oscillatory, Doppler‑shifted branch of the DM (where the frequency in the DM rest frame is $\omega_r = 0$, resulting in a Doppler‑shifted frequency $x = y v_r$) occurs at
\begin{eqnarray}
y^2 -1 = y^2 v_r^2 ~ \Rightarrow ~ y^2 = 1/(1-v_r^2)
\label{Eq:Res}
\end{eqnarray}
and the real oscillation frequency of this mode is approximately $x_r = y v_r = v_r/\sqrt{ 1-v_r^2}$.

\begin{figure}
\includegraphics[width=\linewidth]{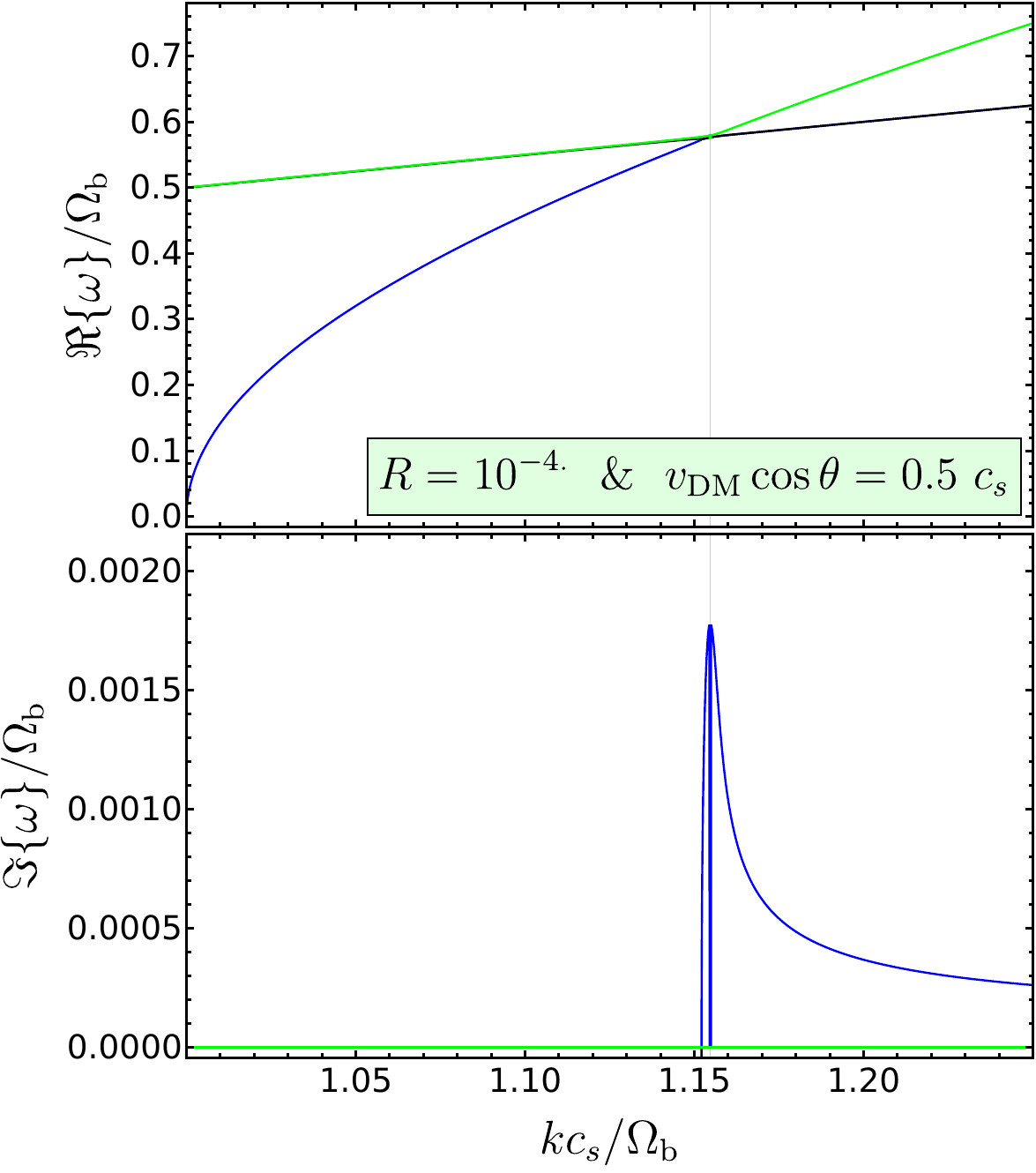} 
\caption{\label{fig:DSsol}%
Solutions for the dispersion relation (Equation ~\eqref{Eq:disp18}) with parameters given in the top-panel inset. Vertical gray lines show the expected resonance scale given by Equation \eqref{Eq:Res}. This show that the streaming-driven resonance has growth rate which is maximized at the resonance wavelength given by 
$k c_s/\OB = 1/\sqrt{1-v_r^2}$.}
\end{figure}

This resonance occurs only if $v_r <1$, i.e., $v_{\rm DM} \cos \theta < c_s$.
In this regime, DM drives the gravitational perturbation to become unstable, with the growth rate given by $\Gamma_{\rm DM} = \Omega_{\rm DM} ~ \Rightarrow \Im[x] = R$.
Therefore, the question of whether such resonance really enhances linear growth can be recast as follows: Can the growth rate exceed that derived from DM itself, and if so, by how much.
In Figure~\ref{fig:DSsol}, we show an example of the case where the resonantly driven instability enhances the growth of the gravitational perturbation:
We present the solution for $R=10^{-4}$ and $v_r = 0.5$ case.

\subsubsection{Growth rate at resonance scale}

The dispersion relation in Equation~\eqref{Eq:disp01} is quadratic and can be easily solved. Below, we show an approximate solution at the streaming resonance scale, i.e, at $y = 1/\sqrt{ 1-v_r^2}$, and we define 
$u_r = v_r/\sqrt{ 1-v_r^2}$. In this case, the dispersion relation can be re-written as
\begin{eqnarray}
-1 = 
\frac{1}{ x^2 - u_r^2 -1}+ 
\frac{R^2}{
(x - u_r)^2}
\label{Eq:disp02}
\end{eqnarray}

The solution of Equation \eqref{Eq:disp02} is complicated; the branch that has the growing solution, i.e, $\Im[x] >0$, is such that 
\begin{eqnarray}
2 x &=&  
- \sqrt{-\frac{2 u_r \left(u_r^2+R^2\right)}{\sqrt{\sqrt[3]{c_2}+c_3}}-\sqrt[3]{c_2}-c_3+3 u_r^2-2 R^2}
\nonumber \\ 
&&
+ \sqrt{\sqrt[3]{c_2}+c_3}+u_r
\end{eqnarray}
Where,
\begin{eqnarray}
c_3 &=& \frac{R^2 \left(-12 u_r^2+R^2-12\right)}{9 \sqrt[3]{c_2}}+u_r^2-\frac{2 R^2}{3}
\\
c_2 &=&
-\frac{2 \sqrt{c_1}}{3 \sqrt{3}}+\frac{4}{3} R^4 u_r^2-2 R^2 u_r^2+\frac{R^6}{27}+\frac{4 R^4}{3}
\\
c_1 &=&
R^{10} u_r^2+8 R^8 u_r^4+15 R^8 u_r^2+16 R^6 u_r^6+12 R^6 u_r^4
\nonumber \\ &&
+12 R^6 u_r^2+27 R^4 u_r^4+R^{10}+8 R^8+16 R^6
\end{eqnarray}

The resonant growth rates are a function of $R$ \& $u_r$. This formula is complicated to work with; however, if we expand this solution for $R\ll1$, we get
\begin{eqnarray}
x =
\frac{(-R)^{2/3}}{ (2 u_r)^{1/3}} +u_r
\quad \Rightarrow \quad
\Im[x] = \frac{\sqrt{3}}{2 \sqrt[3]{2 u_r }} 
 R^{2/3} 
\label{Eq:ResDisp}
\qquad
\end{eqnarray}

\begin{figure}
\includegraphics[width=\linewidth]{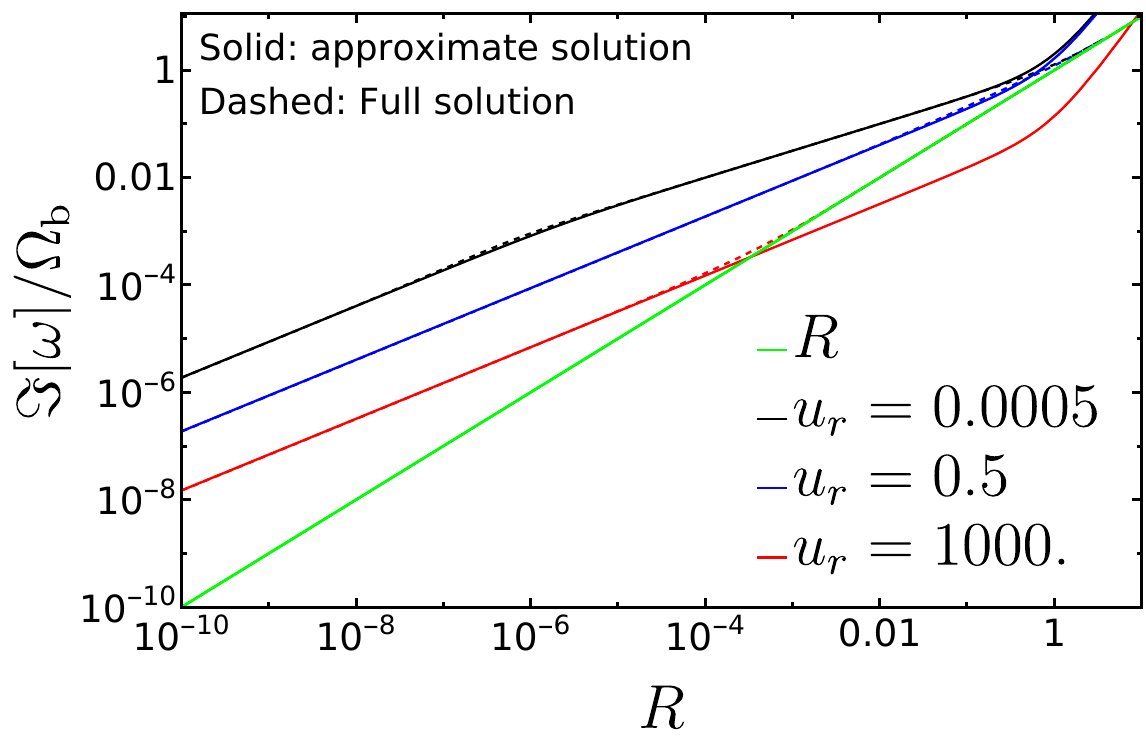}
\caption{\label{fig:Rdep}%
Validating the approximate solution for the resonantly driven growth rate given in Equation \eqref{Eq:GrR}: The dependence of the growth rate on $R$ at various values of $u_r = v_r/\sqrt{1-v_r^2} $.
}
\end{figure}

\begin{figure*}
\centering
\includegraphics[height=5.72cm]{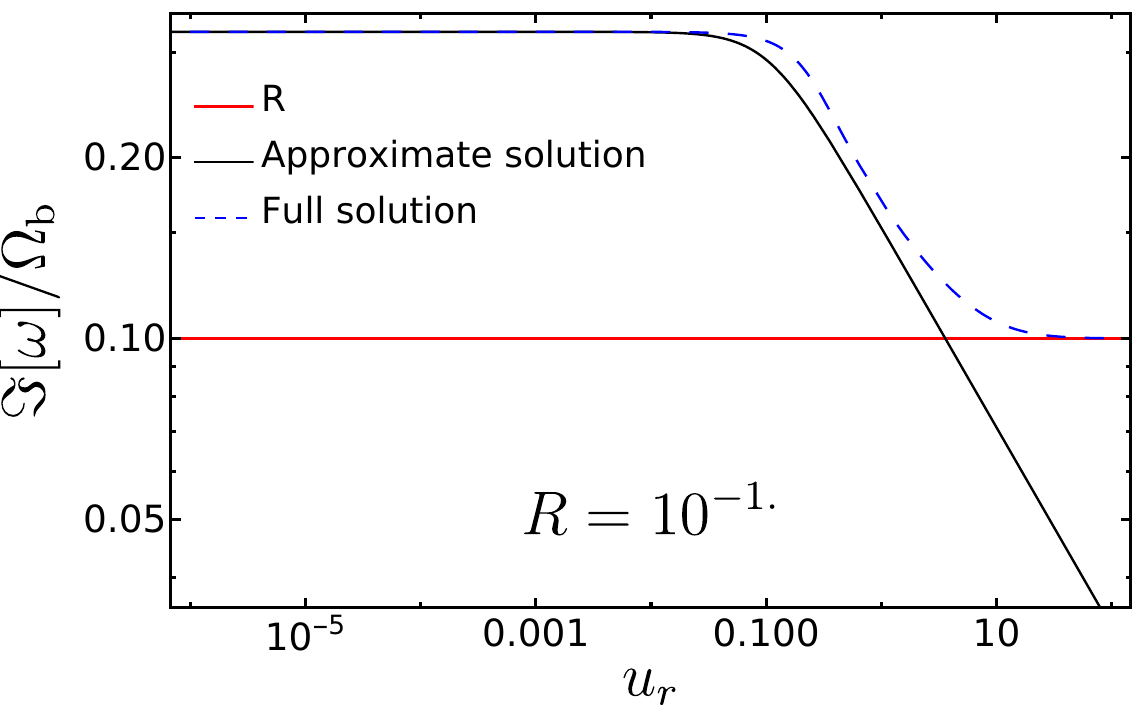}
\includegraphics[height=5.72cm]{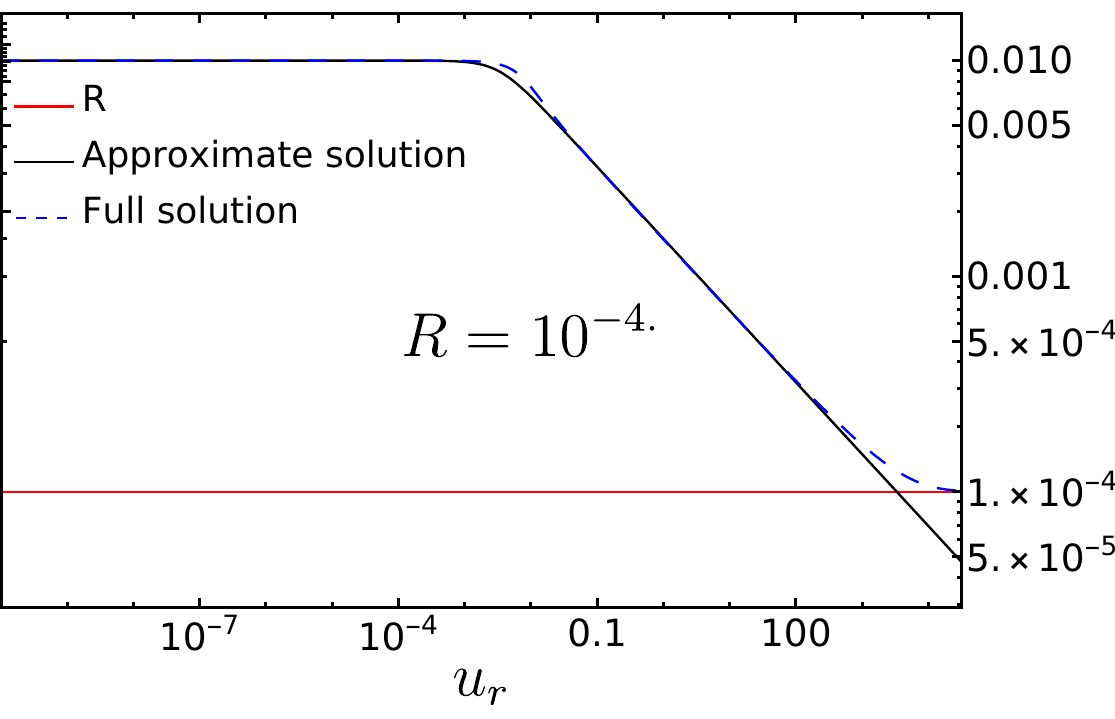}
\caption{\label{fig:urdep}%
Validating the approximate solution for the resonant growth rate given in Equation \eqref{Eq:GrR}: The dependence of the growth rate on $u_r$ at various values of $R$.
}
\end{figure*}

This, as expected, does not agree very well with the imaginary part of the full solution for values of $R$ close to $1$.
However, inspired by this solution, we can construct a numerical formula that agrees much better with the numerical growth rates at resonance. This formula is given by
\begin{eqnarray}
\boxed{
\Gamma_{\rm res} = 
\frac{ \left(R^{8/5}+1\right)}{\sqrt[6]{9 u_r^2+R}} R^{2/3}
}
\label{Eq:GrR}
\end{eqnarray}

To validate this approximation, we compare in Figures \ref{fig:Rdep} and \ref{fig:urdep} the growth rate at the resonant scale (full solution) with the approximation given in Equation~\eqref{Eq:GrR}.
Both Figures show excellent agreement and, more importantly, that the approximation in Equation~\eqref{Eq:GrR} is valid as long as the resonant growth rate is larger compared to that driven by cold DM, $\Gamma_{\rm DM} = \ODM$, i.e., $\Im\{x\} = R$.

Therefore, the growth rate in Equation~\eqref{Eq:GrR} provides an excellent approximation of the true growth rate at the resonance scale for all values of $R \leq 1$ when $\Gamma_{\mathrm{res}} > R$.
It is important to note that this approximate solution, while highly accurate, can yield values slightly below the true growth rate in edge cases, as seen in Figures \ref{fig:Rdep} and \ref{fig:urdep}. Consequently, using this approximation will never overestimate the  growth rates.

\subsection{Impact of DM velocity dispersion on the resonant instability}
\label{app:hdrft}

This would be relevant to understanding the neutrino in the early universe since it acts like hot DM, but their contribution should be only $1\%$ of that of DM. 
So the low value of $R$ is the more relevant case.
In this case, we solve the full dispersion relation given by
\begin{eqnarray}
-1 = 
\frac{\Omega_b^2 }{ \omega^2 - k^2 c_s^2}+ 
\frac{\Omega_{\rm DM}^2}{
(\omega - \vec{k} \cdot \vec{v}_{\rm DM})^2 - k^2 c_{\rm DM}^2}
\label{Eq:dispCw00}
\end{eqnarray}
In dimensionless variables, the dispersion relation is given by
\begin{eqnarray}
-1 = 
\frac{1}{ x^2 - y^2}+ 
\frac{R^2}{
(x - y v_r)^2 - y^2 c_r^2}
\label{Eq:disp04}
\end{eqnarray}

\subsubsection{Resonance Location}

Assuming $R \ll 1$, the location of the fastest growth rates can be found as follows
\begin{eqnarray}
(x - y v_r) = \pm y c_r
&&~~ \rightarrow ~~
x  = y (v_r \pm c_r) = \sqrt{y^2-1}
\nonumber \\
&&~~ \rightarrow ~~
y^2 = \frac{1}{1-(v_r \pm c_r)^2}
\end{eqnarray}

From solving the dispersion relation in this case (restricted to the case $v_r >0$), the growth is nonzero only if $v_r>c_r$ and is maximized  at
\begin{eqnarray}
y^2 = \frac{1}{1-(v_r - c_r)^2}
\end{eqnarray}

To find the fastest growth rate, we define the following two useful quantities at the resonant scale
\begin{eqnarray}
q_r&\equiv& y c_r 
= 
\frac{c_r}{\sqrt{1- (v_r-c_r)^2}} 
\\
u_r &\equiv& y v_r
\Rightarrow 
(u_r-q_r)^2 +1 = y^2 \qquad  ~~~~  
\end{eqnarray}

Therefore, the dispersion relation at the resonant scale can be written as
\begin{eqnarray}
-1 
&=& 
\frac{1}{ x^2 - (u_r-q_r)^2 -1}
+ 
\frac{R^2}{
(x - u_r)^2 - q_r^2}
\label{Eq:disp05}
\end{eqnarray}
In the limit of $c_r =0$, i.e., $q_r=0$, The dispersion relation in Equation \ref{Eq:disp05} is the same as that in Equation \ref{Eq:disp02}.

\subsubsection{Growth rates at resonance scale}

\begin{figure*}[htbp] 
\includegraphics[width=1.0\linewidth]{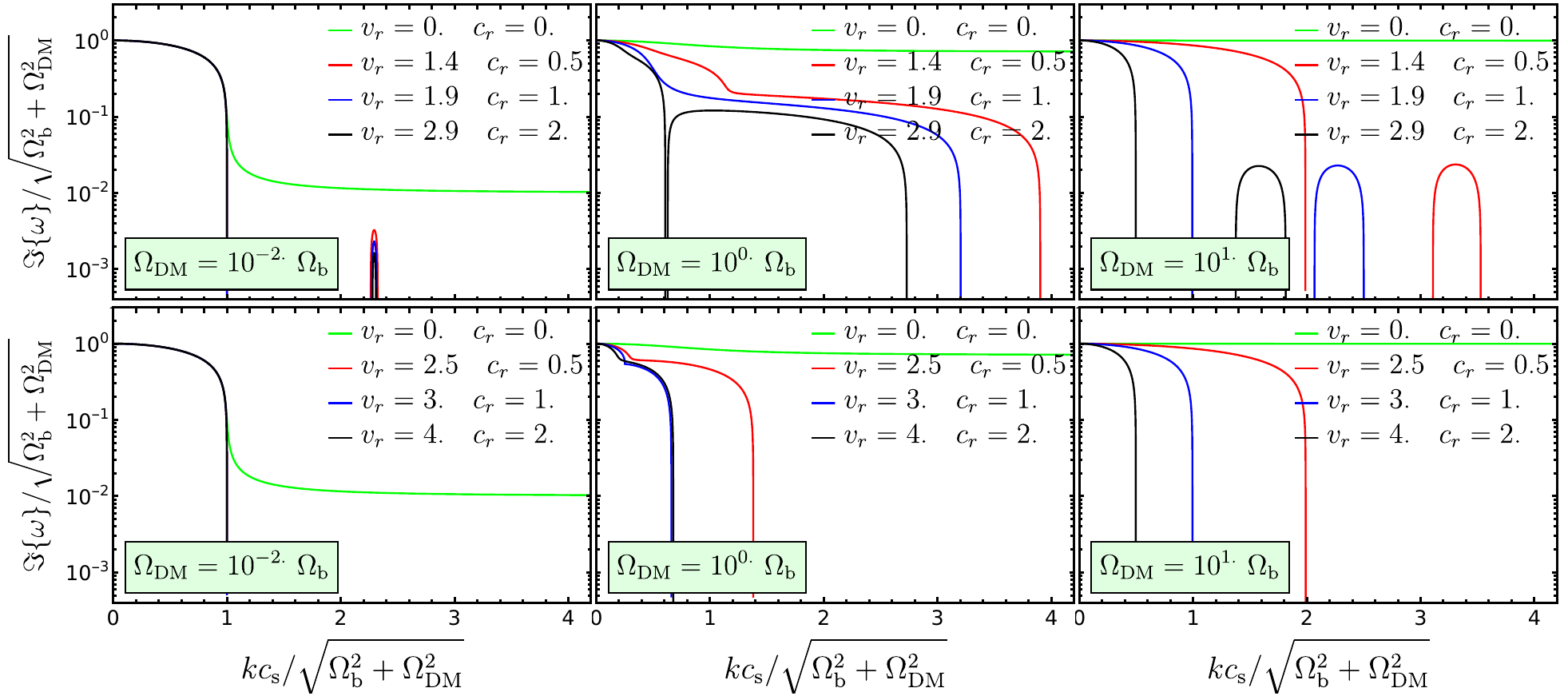}
\caption{\label{fig:res_enhcdm}%
Top panel show resonant instability growth rates ($v_r-c_r=0.9$ for all cases) at various sound speed ratios ($c_r = c_\mathrm{DM}/c_s$, and DM to baryon density ratios, $R = \ODM /\OB $.
Bottom panel shows the case of supersonic relative drift ($v_r-c_r=2$ for all cases).
For comparison, we show for all cases, the growth rates for the cases of cold non-drifting DM are shown in green in all panels.
}
\end{figure*}

Figure~\ref{fig:res_enhcdm} presents numerical solutions to the dispersion relation across various parameter regimes. The top panels show subsonic relative drift, where \(v_r - c_r < 1 \Rightarrow v_{\rm DM} \cos \theta < c_s + c_{\rm DM}\), with the effective sound speed given by the sum of the baryon and dark matter sound speeds. The bottom panels show supersonic cases. For comparison, the cold, non-drifting dark matter case appears in green in all panels.

In the baryon-dominated regime with subsonic drift (top-right panel), the growth rate enhances at the resonance scale. Here, the dark matter sound speed stabilizes all wavelengths shorter than the effective Jeans scale. However, a supersonic relative drift destabilizes sound waves at the resonance scale, and the dark matter sound speed (whether smaller or larger than that of baryons) does not stabilize this resonant instability. In the supersonic drift case (bottom-right panel), no resonance occurs, and consequently, no unstable modes appear at wavelengths shorter than the effective Jeans scale.

When baryons and dark matter have comparable mass densities, the resonant instability significantly enhances growth at wavelengths shorter than those in the supersonic case (compare top and bottom middle panels). Similarly, in the dark matter-dominated regime, subsonic drift produces a resonant instability that counteracts the stabilization otherwise provided by the dark matter sound speed.

\subsection{Non-resonant supersonic stability}

The bottom panel of Figure \ref{fig:res_enhcdm} shows the case where the resonance condition is not fulfilled since $v_\mathrm{DM} \cos \theta > c_\mathrm{DM} + c_s$.
We see that when the baryon density is dominant, this supersonic relative drift does not impact the growth rates of the unstable modes.
When the density of dark matter is comparable to that of baryons (middle lower panel), we can see that the supersonic drift leads to a small suppression of the growth rates of the long wavelength modes, i.e., in this case, wavelengths $5-10$ are longer compared to the effective Jeans wavelength at $k c_c = \sqrt{\ODM^2 + \OB^2}$.

When the DM density is larger compared to that of baryons (right bottom panel), the supersonic drift does not change the growth rates compared to the no drift cases shown as green curves in the bottom panels of Figure \ref{fig:novdrCr}.

\section{Governing equations in an expanding/cosmological spacetime}
\label{sec:lincosm}

The linear equations of the coupled baryon-DM system with relative drift $v_b$ in the baryon rest frame are given by: (Equation 11 of \citet[][]{Tseliakhovich+Hirata_2010})
\begin{eqnarray}
 \partial_t \delta_b &=& - \theta_b 
\quad \& \quad
\partial_t \delta_{\rm DM} =  - \theta_{\rm DM}  + i \frac{\vec{k} \cdot \vec{v}_{\rm DM}}{a(t)}  \delta_{\rm DM}
\\
\partial_t  \theta_b   &=&  
\frac{c_s^2 k^2}{a^2} \delta_b
-\frac{ 3 H^2 }{2} (\Omega_b \delta_b  + \Omega_{\rm DM} \delta_{\rm DM})
- 2 H \theta_{b} 
\\
\partial_t  \theta_{\rm DM}   &=&
i \frac{\vec{k} \cdot \vec{v}_{\rm DM}}{a(t)}  \theta_{\rm DM}
-\frac{ 3 H^2 }{2} (\Omega_b \delta_b  + \Omega_{\rm DM} \delta_{\rm DM})
- 2 H \theta_{\rm DM} 
\nonumber
\\
\end{eqnarray}
Where, $\delta_{b(\rm DM)} (\vec{x},t) \equiv (\rho(\vec{x},t) -\rho_0(t) )/ \rho_0(t)$ is the over-density compared to the cosmologically evolving background density $\rho_0(t)$ for baryon (DM); $a(t)$ is the scale factor, $\theta_{b({\rm DM)}} \equiv \vec{k}\cdot \vec{v}_{1,b({\rm DM)}}/a(t)$, and $H = \dot{a}/a$ is the Hubble parameter.
Since $\theta$ has the same dimension as $H_0$ (the Hubble parameter at the current epoch), we define $t = \theta/H_0$; thus, the equations become:
\begin{eqnarray}
\partial_t \delta_b &=& 
- H_0 t_b  
\nonumber
\\
\partial_t \delta_{\rm DM}  &=& 
- H_0  t_{\rm DM}
+ i \frac{\vec{k} \cdot \vec{v}_{\rm DM}}{a(t)}  \delta_{\rm DM}
\nonumber
\\
 \partial_t  t_b   &=&  
\frac{c_s^2 k^2}{a^2 H_0  } \delta_b
-\frac{ 3 H^2 }{2 H_0 } (\Omega_b \delta_b  + \Omega_{\rm DM} \delta_{\rm DM})
- 2 H  t_{b} 
\nonumber
\\
\partial_t  t_{\rm DM}   &=&
i \frac{\vec{k} \cdot \vec{v}_{\rm DM}}{a }   t_{\rm DM}
-\frac{ 3 H^2 }{2 H_0  } (\Omega_b \delta_b  + \Omega_{\rm DM} \delta_{\rm DM})
- 2 H t_{\rm DM} 
\nonumber
\\
\end{eqnarray}

To simplify, we replace the time derivatives with a derivative with respect to the scale factor $a$, i.e., $\partial_t = a H \partial_a$; thus, the equations read.
\begin{eqnarray}
a \partial_a \delta_b &=& 
- \frac{t_b}{H /H_0}   
~~~ \& ~
a \partial_a \delta_{\rm DM} =  
- \frac{t_{\rm DM}}{H /H_0}   
+ i \frac{\vec{k} \cdot \vec{v}_{\rm DM}}{a H }  \delta_{\rm DM}
\nonumber
\\
 a \partial_a t_b   &=&  
\frac{c_s^2 k^2}{a^2 H H_0  } \delta_b
-\frac{ 3 H_0 }{2 a^3  H } (\Omega_{b,0} \delta_b  + \Omega_{\rm DM,0} \delta_{\rm DM})
- 2   t_{b} 
\nonumber
\\
a  \partial_a  t_{\rm DM}   &=&
i \frac{\vec{k} \cdot \vec{v}_{\rm DM}}{a H}   t_{\rm DM}
 - 2 t_{\rm DM} 
\nonumber
\\
&& -\frac{ 3 H_0/H}{2a^3} (\Omega_{b,0} \delta_b  + \Omega_{\rm DM,0} \delta_{\rm DM}) 
\label{eq:lincosm}
\end{eqnarray}
Where, $H = H_0 \sqrt{ \Omega_{b,0}/a^3  +  \Omega_{\rm DM,0}/a^3  + \Omega_{\Lambda,0} } $.
The cosmology we adopt here is:
$ \Omega_{b,0} = 0.044$, 
$\Omega_{\rm DM,0} =  0.226$, 
$\Omega_{\Lambda,0} = 0.73 $, and
$ H_0 = 71 $km/sec/Mpc.
Since $c_s$, $H$, and $v_{\rm DM}$ are all functions of time, proceeding by taking the Laplace transform with respect to     time leads to complicated convolutions.
Thus, the solutions are typically found by solving these equations numerically.

Therefore, to proceed, we need an expression for the time evolving baryon sound speed. The baryon temperature $T_{\rm b}$ and sound speed are given by~\citep{Tseliakhovich+Hirata_2010}
\begin{eqnarray}
\label{eq:Tb}
   T_{\rm b} &=&
   \frac{ T_{\rm cmb} }{a} 
   \left( 
   1 + \frac{ a/a_1}{ 1 + (a_2/a)^{3/2} }
   \right)^{-1}
\nonumber \\  c_s &=& \sqrt{
    \frac{\Gamma k_B T_b}{ \mu m_\mathrm{p}}
   }, 
\end{eqnarray}
where CMB temperature $T_{\rm cmb} = 2.7$ K, $a_1 = 1/119$, $a_2 = 1/115$, mean molecular weight $\mu = 1.22$, the adiabatic index $\Gamma = 5/3$, $k_B$ is the Boltzmann constant, and the cosmological scale factor $a = 1/(1 + z)$, where $z$ is the cosmological redshift. At $z=1000$, the relative drift is such that $v_{\rm DM} = 5 ~c_s$ in the baryon rest frame~\citep{Tseliakhovich+Hirata_2010}.
For a more detailed derivation of the baryon temperature evolution and the effects of spatial fluctuations in the sound speed on small-scale perturbations~\citep[see, e.g.,][]{Naoz+2005}.

\begin{figure}
\includegraphics[width=\linewidth]{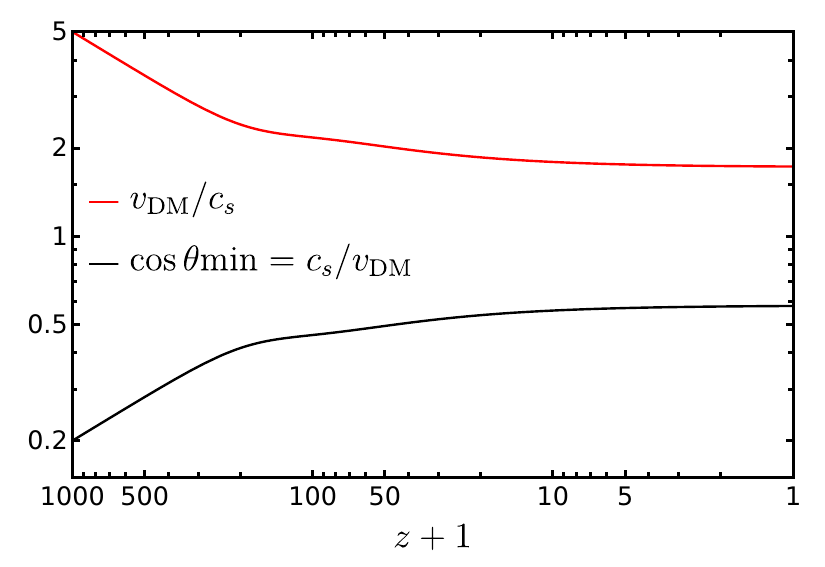}
\caption{\label{fig:cs}%
The cosmological evolution of the DM--baryon relative drift, normalized to the evolving baryon sound speed, is shown in red. The black curve indicates the minimum angle \(\theta_{\mathrm{min}}\) between the wave mode and the DM streaming direction for which the projected speed equals the sound speed, i.e., \(v_r = v_{\mathrm{DM}}\cos\theta / c_s = 1\). All wave modes with angles \(\theta \in [\theta_{\mathrm{min}}, \pi/2)\) can be resonantly driven.
That is while the range for resonance at high redshift is small, i.e., $\cos \theta < 0.2$ the decreasing values of $v_{\rm DM}/c_s$ at lower $z$ means that the  window for wave modes that can be resonantly excited by the relative drift is wider at higher redshift, $\cos \theta < 0.5$ at $z < 100$.
}
\end{figure}

In Figure~\ref{fig:cs}, we show the evolution of the DM–baryon relative speed, normalized to the evolving baryon sound speed. For the resonant instability to be possible, the projected DM speed must be subsonic. This condition defines a minimum angle $\theta_{\mathrm{min}}$ such that $v_{\mathrm{DM}}\cos\theta_{\mathrm{min}} = c_s$; all modes with $\theta > \theta_{\mathrm{min}}$ have subsonic projected drifts and can be resonantly driven. However, for $\theta = \pi/2$, the projected drift is exactly zero, and no resonant amplification occurs.

At high redshift ($z \approx 1000$),     $v_{\mathrm{DM}}/c_s$ is about 5, and thus $\theta_{\mathrm{min}}$ is about $ 0.87 \times \pi/2$; consequently, only modes almost perpendicular to the DM stream are subsonic, giving a narrow resonant window (e.g., $\cos\theta < 0.2$ at $z\sim1000$). 
As the universe expands, $v_{\mathrm{DM}}$ decays as $1/a$ while the baryon sound speed decreases more slowly, causing $v_{\mathrm{DM}}/c_s$ to decrease. This reduction in the relative speed broadens the range of angles for which the projected drift is subsonic. At $z \approx 100$, the condition becomes $\cos\theta < 0.5$, meaning that a much wider range of wave modes is resonantly excited.
Thus, the resonant instability is  relevant across a wide range of redshifts, with the angular window widening as the universe ages.

\section{Dispersion in elastic solids}
\label{app:elasticsolids}

As discussed in Section~\ref{sec:direct}, the dispersion relation in elastic solids is analogous to that in a thermal gas, albeit with a propagation speed that depends on the Lam\'e constants. We provide here a brief derivation, closely following the presentation in Chapter~5 of \citep{Billingham+Waves}.

In elastic solids, the Euler equations (Equation~\eqref{Eq:momentum}) are not strictly applicable. However, when studying the propagation of small-amplitude (linear) waves in ideal elastic bodies, the derivative of the displacement vector $\vec{u} \equiv \vec{X} - \vec{x}$ replaces the velocity in the Euler equation, where $\vec{X}$ and $\vec{x}$ are the positions before and after the displacement, respectively.
The time evolution of the displacement vector is governed by
\begin{equation}
    \rho_0 \frac{\partial^2 u_i}{\partial t^2}
    = \frac{\partial \sigma_{ij}}{\partial x_j} + a_i,
\end{equation}
where $\rho_0$ is the unperturbed mass density of the elastic solid, $u_i$ denotes the $i$-th component of $\vec{u}$ ($i \in \{x,y,z\}$), $\vec{a}$ is the acceleration due to self-gravity, and the stress tensor $\sigma_{ij}$ is given by
\begin{equation}
    \sigma_{ij} = \lambda e_{kk} \delta_{ij} + 2\mu e_{ij},
\end{equation}
with the strain tensor $e_{ij} \equiv (u_{i,j} + u_{j,i})/2$, and $\lambda$ and $\mu$ are the Lam\'e constants.

In vector form, the displacement evolution is given by
\begin{equation}
    \rho_0 \frac{\partial^2 \vec{u}}{\partial t^2}
    = (\lambda + \mu) \nabla (\nabla \cdot \vec{u}) + \mu \nabla^2 \vec{u}
    + \vec{a}.
\end{equation}

Introducing the dilation $\Delta \equiv \nabla \cdot \vec{u}$, this can be recast as
\begin{equation}
    \rho_0 \partial_t^2 \Delta = \rho_0 c_1^2 \nabla^2 \Delta + \nabla \cdot \vec{a},
\end{equation}
which has the same form as the divergence of Equation~\eqref{Eq:velocity}, with the propagation speed of vibrational (sound) waves in this case is given by
\begin{equation}
    c_1 = \sqrt{\frac{\lambda + 2\mu}{\rho_0}}.
\end{equation}

The treatment of the self-gravity term $\vec{a}$ follows exactly the same procedure as in Appendix~\ref{app:disper}. This demonstrates that the linear response of an elastic solid can be obtained simply by replacing the sound speed $c_s$ of the fluid with the vibrational speed $c_1$ of the solid, which depends on the Lam\'e constants.

\end{document}